\documentclass[journal]{IEEEtran}
\usepackage{xspace}
\usepackage{enumerate}
\usepackage{url}

\usepackage[ruled]{algorithm}
\usepackage[noend]{algpseudocode}
\usepackage{multirow}
\usepackage[pdftex]{graphicx}
\usepackage{amsmath}
\usepackage{amssymb}
\usepackage{cleveref}
\usepackage{xcolor}
\usepackage{graphicx}
\usepackage{listings}

\def\oursystem{AppAngio\xspace}
\def\ie{\textit{i.e.}\xspace}
\def\etc{\textit{etc.}\xspace}
\def\eg{\textit{e.g.}\xspace}
\def\code#1{\texttt{#1}}

\newcommand{\tabincell}[2]{\begin{tabular}{@{}#1@{}}#2\end{tabular}}

\def\true{\bullet}
\def\truenegative{\blacksquare}
\def\falsepositive{\circ}
\def\falsenegative{\square}

\begin{document}

\title{AppAngio: Revealing Contextual Information of Android App Behaviors by API-Level Audit Logs}

\author{Zhaoyi~Meng,
        Yan~Xiong,
        Wenchao~Huang,
        Fuyou~Miao,
        Jianmeng~Huang
\thanks{Zhaoyi Meng, Yan Xiong, Wenchao Huang, Fuyou Miao, Jianmeng Huang are with the School of Computer Science and Technology, University
of Science and Technology of China, Hefei, 230027, China. Corresponding author: Wenchao Huang. (e-mail: \{mzy516, yxiong, huangwc, mfy\}@ustc.edu.cn, mengh@mail.ustc.edu.cn)}}

\maketitle

\begin{abstract}
Android users are now suffering severe threats from unwanted behaviors of various apps.
The analysis of apps' audit logs is one of the essential methods for the security analysts of various companies to unveil the underlying maliciousness within apps.
We propose and implement \textit{\oursystem}, a novel system that reveals contextual information in Android app behaviors by API-level audit logs.
Our goal is to help security analysts understand how the target apps worked and facilitate the identification of the maliciousness within apps.
The key module of \oursystem is identifying the path matched with the logs on the app's control-flow graphs~(CFGs).
The challenge, however, is that the limited-quantity logs may incur high computational complexity in the log matching, where there are a large number of candidates caused by the coupling relation of successive logs. 
To address the challenge, we propose a divide and conquer strategy that precisely positions the nodes matched with log records on the corresponding CFGs and connects the nodes with as few backtracks as possible.
Our experiments show that \oursystem reveals contextual information of behaviors in real-world apps.
Moreover, the revealed results assist the analysts in identifying the maliciousness of app behaviors and complement existing analysis schemes.
Meanwhile, \oursystem incurs negligible performance overhead on the real device in the experiments.
\end{abstract}


\IEEEpeerreviewmaketitle

\section{Introduction}
\label{sec:introduction}

\IEEEPARstart{W}{ith} the growing popularity of the Android platform, the threats from unwanted behaviors of various apps, including malware and other potentially harmful apps, have become more serious~\cite{pan2017dark}.
The apps may leak users' private information without consent, root users' devices silently, send premium SMS stealthily, \etc, which have already affected the dependability of the Android app ecosystem~\cite{zhang2015android}.

To ensure the security and privacy of Android users, security analysts of various companies perform postmortem analysis based on the specialized audit logs.
Specifically, some device manufacturers have collected log data of users from real devices to diagnose potential Android attacks and generate improved security policies \cite{wang2015easeandroid,Xiaomi,GoogleCorp.}.
Moreover, market operators have adopted different techniques that perform app auditing based on the logs from the configured
emulators and remove the detected malware from the regulated markets \cite{DBLP:conf/sp/XiaGLQL15,isohara2011kernel,zhou2012hey}.
In the scenarios, the more information the analysts learn based on the logs, the more likely they are to unveil underlying maliciousness.
Therefore, the precise and complete reconstruction of app behaviors according to the logs is one of the most crucial problems that the analysts concern about.

Many state-of-the-art techniques \cite{isohara2011kernel,yan2012droidscope,DBLP:conf/ndss/TamKFC15,yuan2017droidforensics,DBLP:conf/ccs/ZhangYXYGNWZ13} have been proposed to assist analysts in reconstructing Android app behaviors based on runtime logs.
For example, DroidScope~\cite{yan2012droidscope} reconstructs both the OS-level and Java-level semantics by instrumenting the virtual machine.
CopperDroid~\cite{DBLP:conf/ndss/TamKFC15} leverages system call-related information to reconstruct app behaviors automatically.
DroidForensics~\cite{yuan2017droidforensics} captures multi-layer forensic logs from the application level, Binder level, and system-call level to reconstruct Android attacks.
These proposed schemes have achieved the behavior reconstruction of many real-world apps, but there still exists the problem of the lack of contextual information within the reconstructed behaviors.
Specifically, it is impractical to deploy runtime mechanisms with heavyweight computations (\eg, recording too much information from the system or precisely tracking runtime information flows~\cite{enck2014taintdroid}) for behavior reconstruction due to the resource-constrained feature of Android devices or emulators.
Instead, OS developers have to log some key points (\eg, sensitive Android API calls or system calls~\cite{DBLP:conf/ndss/TamKFC15,yuan2017droidforensics}) to rebuild coarse-grained behaviors.
However, contextual information (\eg, guarding conditions of sensitive actions~\cite{7546513,yang2015appcontext}, obfuscated strings~\cite{rasthofer2016harvesting}, malicious payloads~\cite{li2017understanding}) is unavailable in the app behaviors reconstructed by the logging strategy mentioned above.
The information is valuable evidence to disclose the intentions of app behaviors.

We propose and implement \textit{\oursystem}\footnote[1]{AppAngio stands for \underline{App} \underline{Angio}gram.} \cite{meng2019divide}, a novel system that reveals contextual information of Android app behaviors by API-level audit logs.
\oursystem gathers runtime logs about the target app from the real Android device or emulator and then extracts the path matched with the logs from the app's CFGs offline.
The contextual information profiled from the path enriches the semantics of the behaviors reconstructed via the schemes mentioned above.
Different from the above-mentioned schemes that collect the low-level information from the Android OS, \oursystem only requires to log Android API calls for the contextual reveal, so the imposed performance overhead is negligible.
The goal of \oursystem is to help security analysts understand how the target app worked and facilitate the identification of maliciousness within apps, instead of directly reporting malicious app behaviors.
Moreover, the result of \oursystem can be used to complement existing analysis schemes (\eg, information-flow analysis~\cite{DBLP:conf/pldi/ArztRFBBKTOM14}, behavior reconstruction~\cite{DBLP:conf/ccs/ZhangYXYGNWZ13}).
To the best of our knowledge, \oursystem is the first to reveal contextual information in Android app behaviors by combining audit logs with apps' bytecode.

The major challenge of implementing \oursystem is the high computational complexity of the log matching.
It is caused by the coupling relation that is when a node is matched, its successors are the candidates for matching subsequent logs.
There may be multiple nodes being candidates for matching a log record involving reflection~\cite{li2016droidra}, inter-component communication (ICC)~\cite{octeau2016combining}, transitions in the Android lifecycle~\cite{DBLP:conf/pldi/ArztRFBBKTOM14} and other mechanisms.
For example, unresolved reflective calls may hide the call hierarchies of different methods with \texttt{invoke()} calls, which makes that the candidates of the calls are undecided; imprecise ICC links and uncertain transitions in the Android lifecycle may incur that the branches that are not executed at runtime are regarded as candidate paths.
What's worse, to evade malware detection, Android attackers may abuse the mechanisms, which makes the number of possible candidate paths increase exponentially.

To the best of our knowledge, the existing program analysis techniques can solve a part of the aforementioned problems respectively, but it is sophisticated to effectively manage different techniques for solving the various problems in practice.
For example, Harvester \cite{rasthofer2016harvesting} can extract the runtime values of reflective calls, but it does not support slices that span multiple Android components.
IC3 \cite{DBLP:conf/icse/OcteauLDJM15} correctly models the ICC links that can be resolved statically.
Some dynamic analysis tools~\cite{enck2014taintdroid,DBLP:conf/codaspy/RastogiCE13} may extract the required information from the OS.
Because of their differences in the usage scenarios and the implementation approaches, it is challenging to combine them effectively for reaching expected results.
For the log-based techniques, Sherlog \cite{yuan2010sherlog} and lprof \cite{zhao2014lprof} that combine log messages and log printing statements in code to reconstruct the execution sequence do not handle the anti-analysis techniques (\eg, code obfuscation or string encryption~\cite{rasthofer2016harvesting}) on statements.
A straightforward method is to record enough logs to distinguish each branch path, but the imposed performance and space overhead is considerable.

We propose a divide and conquer strategy to address the challenge.
The computational complexity is reduced by dividing the problem of the log matching into multiple independent subproblems.
In each subproblem, we leverage a small amount of information from the runtime call stack to decompose the coupling relation within the log matching and circumvent a part of the defects of static analysis in the graph building.
The information helps locate \textit{log points}\footnote[2]{In this paper, we use the term \textit{log point} to refer to an invocation statement that is used to call a logged API in app code.} individually and avoid backtracking as much as possible during the log matching, so our strategy achieves a precise and efficient path exploration.

As a result, \oursystem outputs the matched path with sufficient contextual information that has wide practical usage.
Specifically, security analysts can leverage various methods (\eg, manual checking, API usage analysis, machine learning-based analysis) to extract concerned evidence for assessing the maliciousness of the app’s behaviors.
For instance, the path can help the analysts generate signatures or extract features automatically for machine learning-based schemes identifying malicious app behaviors. 
The analysts can also achieve a precise information-flow tracking on the path.

Our main contributions are summarized as follows:
\begin{itemize}

\item We propose and implement \oursystem, a novel system to reveal contextual information of Android app behaviors based on API-level logs, and meanwhile imposing negligible performance overhead.

\item We propose a divide and conquer strategy to achieve a precise and efficient log matching.

\item Our evaluations on the open-source benchmarks \cite{droidbench,iccbench} and real-world apps \cite{allix2016androzoo,DBLP:conf/sp/ZhouJ12,virusshare} validate the effectiveness of \oursystem.
\end{itemize}

The rest of the paper is organized as follows: 
\Cref{sec:overview} introduces the motivating examples and \oursystem's architecture.
\Cref{sec:log&app_analysis} explains the collection and filtering of logs.
\Cref{sec:divide-and-conquer} illustrates the details of the log matching.
In \Cref{sec:evaluation} presents experimental results on \oursystem.
\Cref{sec:limitations} discusses the limitations and future work.
\Cref{sec:related_work} shows the related work, and we conclude in \Cref{sec:conclusions}.

\section{Motivating Examples \& Architecture Overview}
\label{sec:overview}

\begin{figure}[t]
    \centering 
    \includegraphics[width=0.85\linewidth]{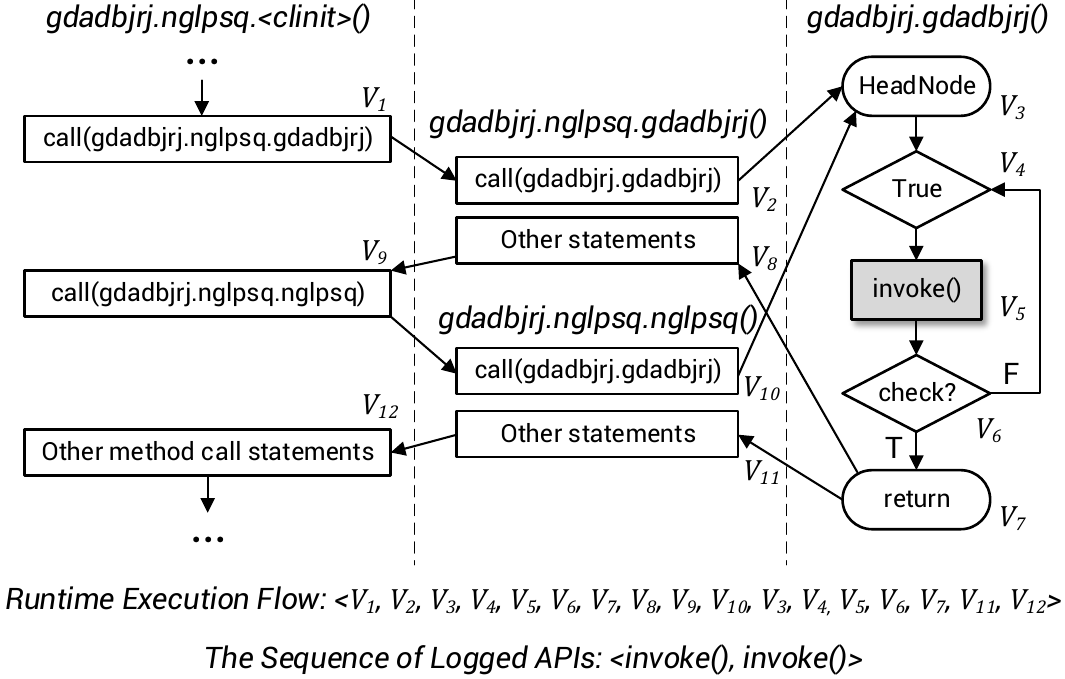}
    \caption{An example for explaining the computational complexity of the log matching for reflection.}
    \label{fig:problemStatement}
\end{figure}

\begin{figure}[t]
    \centering 
    \includegraphics[width=0.9\linewidth]{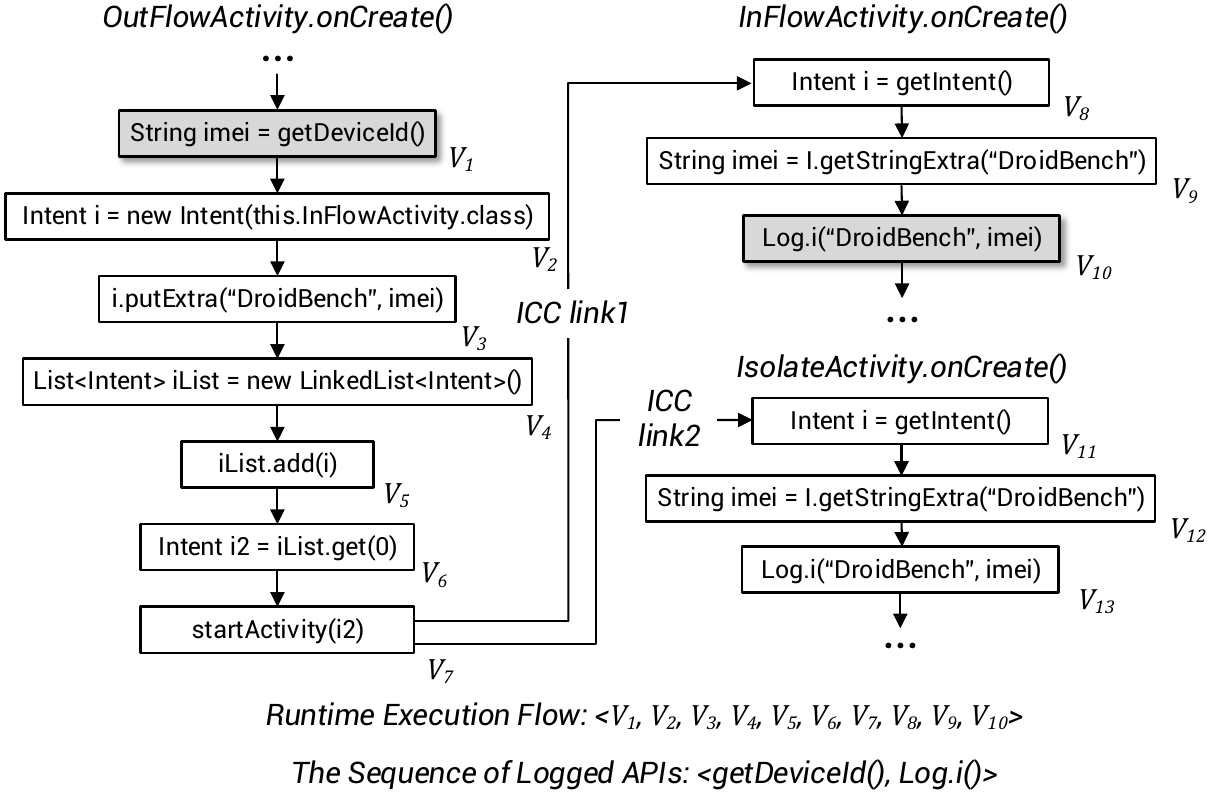}
    \caption{An example for explaining the computational complexity of the log matching for ICC.}
    \label{fig:Icc}
\end{figure}

In this section, we first choose the simplified CFGs in \Cref{fig:problemStatement} and \Cref{fig:Icc} extracted from two representative real-world apps to illustrate the problem and the technical challenge in implementing the log matching algorithm and how the algorithm helps analysts understand app behaviors.
We then describe the architecture of \oursystem.

\begin{table}[]
\caption{The changes of methods in the call stack when the app executes along the runtime execution flow in \Cref{fig:problemStatement}.}
\label{callstack}
\resizebox{\linewidth}{!}{
\begin{tabular}{|l|l|l|}
\hline 
\textbf{\#} & \textbf{Nodes} & \textbf{The Sequence of Methods in the Runtime Call Stack} \\
\hline                                                                                  
1 & \textit{V$_{1}$}   & ..., gdadbjrj.nglpsq.\textless{}clinit\textgreater{}                                           \\
\hline 
2 & \textit{V$_{2}$}   & ..., gdadbjrj.nglpsq.\textless{}clinit\textgreater{}, gdadbjrj.nglpsq.gdadbjrj                    \\
\hline 
3 & \textit{V$_{3}$}   & ..., gdadbjrj.nglpsq.\textless{}clinit\textgreater{}, gdadbjrj.nglpsq.gdadbjrj, gdadbjrj.gdadbjrj \\
\hline 
4 & \textit{V$_{4}$}   & ..., gdadbjrj.nglpsq.\textless{}clinit\textgreater{}, gdadbjrj.nglpsq.gdadbjrj, gdadbjrj.gdadbjrj \\
\hline 
5 & \textit{V$_{5}$}   & ..., gdadbjrj.nglpsq.\textless{}clinit\textgreater{}, gdadbjrj.nglpsq.gdadbjrj, gdadbjrj.gdadbjrj \\
\hline 
6 & \textit{V$_{6}$}   & ..., gdadbjrj.nglpsq.\textless{}clinit\textgreater{}, gdadbjrj.nglpsq.gdadbjrj, gdadbjrj.gdadbjrj \\
\hline 
7 & \textit{V$_{7}$}   & ..., gdadbjrj.nglpsq.\textless{}clinit\textgreater{}, gdadbjrj.nglpsq.gdadbjrj, gdadbjrj.gdadbjrj \\
\hline 
8 & \textit{V$_{8}$}   & ..., gdadbjrj.nglpsq.\textless{}clinit\textgreater{}, gdadbjrj.nglpsq.gdadbjrj                    \\
\hline 
9 & \textit{V$_{9}$}   & ..., gdadbjrj.nglpsq.\textless{}clinit\textgreater{}                                            \\
\hline 
10 & \textit{V$_{10}$}  & ..., gdadbjrj.nglpsq.\textless{}clinit\textgreater{}, gdadbjrj.nglpsq.nglpsq                      \\
\hline 
11 & \textit{V$_{3}$}   & ..., gdadbjrj.nglpsq.\textless{}clinit\textgreater{}, gdadbjrj.nglpsq.nglpsq, gdadbjrj.gdadbjrj   \\
\hline 
12 & \textit{V$_{4}$}   & ..., gdadbjrj.nglpsq.\textless{}clinit\textgreater{}, gdadbjrj.nglpsq.nglpsq, gdadbjrj.gdadbjrj   \\
\hline 
13 & \textit{V$_{5}$}   & ..., gdadbjrj.nglpsq.\textless{}clinit\textgreater{}, gdadbjrj.nglpsq.nglpsq, gdadbjrj.gdadbjrj   \\
\hline 
14 & \textit{V$_{6}$}   & ..., gdadbjrj.nglpsq.\textless{}clinit\textgreater{}, gdadbjrj.nglpsq.nglpsq, gdadbjrj.gdadbjrj  \\
\hline 
15 & \textit{V$_{7}$}   & ..., gdadbjrj.nglpsq.\textless{}clinit\textgreater{}, gdadbjrj.nglpsq.nglpsq, gdadbjrj.gdadbjrj   \\
\hline 
16 & \textit{V$_{11}$}  & ..., gdadbjrj.nglpsq.\textless{}clinit\textgreater{}, gdadbjrj.nglpsq.nglpsq                      \\
\hline 
17 & \textit{V$_{12}$}  & ..., gdadbjrj.nglpsq.\textless{}clinit\textgreater{}             \\                               
\hline 
\end{tabular}
}
\end{table}

\subsection{Problem Statement \& Technical Challenge} \label{problemstatement}

The problem of the log matching is how to find the matched path with as few backtracks as possible.
The computation complexity increases as the number of backtracks increases.
Straightforward algorithms (\eg, the backtracking search) that match the signatures of the logged APIs with the APIs represented by nodes of CFGs do not apply to the problem.

We first explain the computational complexity of the log matching with reflection by \Cref{fig:problemStatement}.
It is a CFG snippet of a real-world app named \textit{ynqgas.mqbgseos}, where the API named \texttt{invoke()} is logged twice when the app runs as the depicted execution flow.
In experiments, we perform the log matching by the backtracking search and then observe that the visit order of nodes may be inconsistent with the runtime execution flow.
Specifically, the inconsistent order is $\langle$\textit{V$_{1}$}, \textit{V$_{2}$}, \textit{V$_{3}$}, \textit{V$_{4}$}, \textit{V$_{5}$}, \textit{V$_{6}$}, \textit{V$_{4}$}, \textit{V$_{5}$}, \textit{V$_{6}$}$\rangle$.
In this case, the algorithm needs to perform a backtrack to find the correct path.
Actually, most logged APIs of this app are \texttt{invoke()}.
Considering that the straightforward algorithms cannot distinguish these log points, they may perform a large number of backtracks to achieve the log matching, which causes a high computational complexity.

We then use \Cref{fig:Icc} to explain the computational complexity of the log matching with ICC.
It is extracted from the app named \textit{ActivityCommunication6} in DroidBench~\cite{droidbench}.
The app runs along the execution flow at the bottom of the figure, where an \texttt{Intent} is passed through a linked-list along $\langle$\textit{V$_{4}$}, \textit{V$_{5}$}, \textit{V$_{6}$}, \textit{V$_{7}$}$\rangle$ in the method named \texttt{OutFlowActivity.onCreate()}.
Due to the limited capabilities of static analysis tools (\eg, IccTA \cite{DBLP:conf/icse/0029BBKTARBOM15}) in tracking the \texttt{Intent} through a list operation statically, two ICC links are built (\ie, \texttt{ICC link1} and \texttt{ICC link2}), where the former is correct but the latter is redundant.
In this circumstance, the straightforward algorithms (\eg, the backtracking search) may explore along $\langle$\textit{V$_{1}$}, \textit{V$_{2}$}, \textit{V$_{3}$}, \textit{V$_{4}$}, \textit{V$_{5}$}, \textit{V$_{6}$}, \textit{V$_{7}$}, \textit{V$_{11}$}, \textit{V$_{12}$}, \textit{V$_{13}$}$\rangle$ mistakenly and then perform backtracks to find the correct path.
If there are multiple imprecise ICC links in the graph, the computational complexity of the algorithms in achieving the log matching may increase exponentially.

In practice, developers (\eg, attackers) may use the aforementioned mechanisms in combination with other mechanisms that are hard to be handled by managing existing techniques~\cite{DBLP:conf/ndss/TamKFC15,enck2014taintdroid,DBLP:conf/pldi/ArztRFBBKTOM14}, \eg, uncertain transitions in the Android lifecycle \cite{DBLP:conf/pldi/ArztRFBBKTOM14}, junk code insertion \cite{rastogi2013droidchameleon}, which incurs that the computational complexity increases to a greater extent.

\subsection{Our Motivation}

We use the logging information from the runtime call stack to locate the log points individually and avoid backtracking as much as possible.
For the case in \Cref{fig:problemStatement}, we position each log point based on the method call sequence for calling the method that contains the log point.
Here, the method call sequence is logged from the runtime call stack.
As listed in \Cref{callstack}, the APIs named \texttt{invoke()} in \textit{V$_{5}$} at the fifth row and the thirteenth row are invoked along different method call sequences.
Specifically, the method call sequence for invoking the \texttt{invoke()} for the first time is listed in the fifth row of \Cref{callstack}.
To find the log point, we need to orderly visit the listed methods (\ie, searching from \textit{V$_{1}$} to \textit{V$_{5}$} in \Cref{fig:problemStatement}).
To continue finding the next log point of the \texttt{invoke()}, we need to search back to \texttt{gdadbjrj.nglpsq.<clinit>()} along the node sequence $\langle$\textit{V$_{5}$}, \textit{V$_{6}$}, \textit{V$_{7}$}, \textit{V$_{8}$}, \textit{V$_{9}$}$\rangle$ and then orderly visit the methods listed in the thirteenth row of \Cref{callstack} (\ie, searching from \textit{V$_{9}$} to \textit{V$_{5}$} in \Cref{fig:problemStatement}).

For the case in \Cref{fig:Icc}, we decide the ICC link covered at runtime according to the recorded caller method of \texttt{Log.i()}.
Specifically, the caller method logged at runtime is \texttt{InFlowActivity.onCreate()}, so we obtain that \texttt{ICC link1} is the correct path for the log matching and \texttt{ICC link2} needs to be pruned.

\begin{figure}[t]
        \begin{center}
        \centering

        \includegraphics[width=0.9\linewidth]{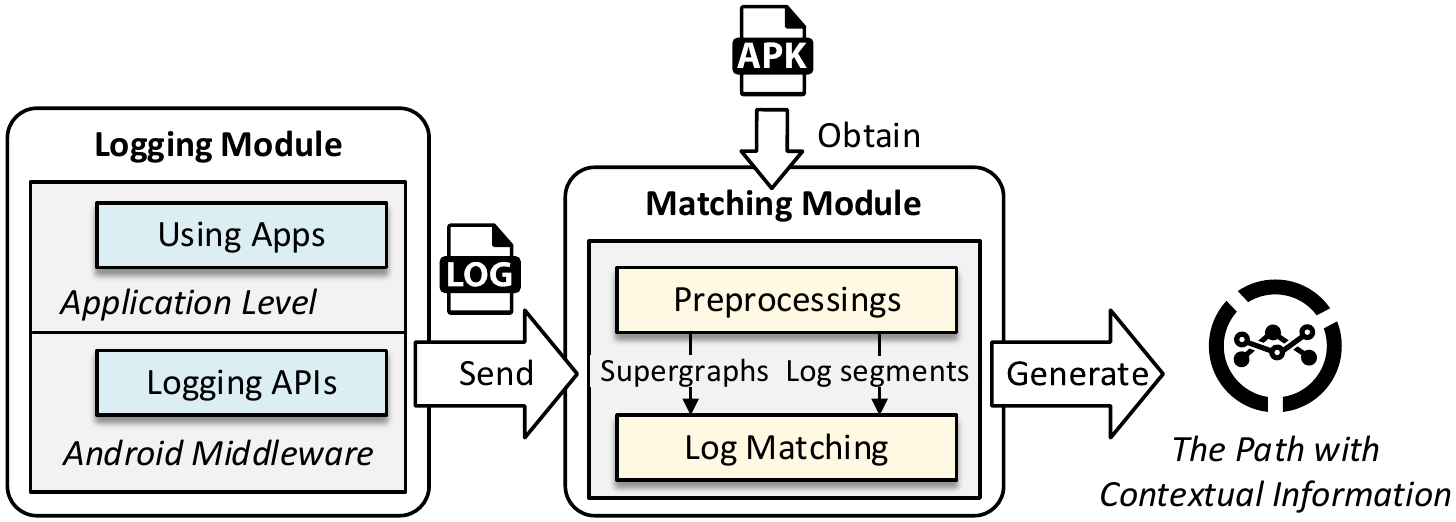}
        \caption{Architecture of \oursystem.}
        \label{fig:overview}
        \end{center}
\end{figure}

\subsection{Architecture of \oursystem}
We therefore design \oursystem that executes the API-level logging and helps analysts identify the path matched with the logs on the target app's CFG.
\Cref{fig:overview} depicts the overall architecture of \oursystem that contains two modules.

\noindent\textbf{Logging Module.} The module is deployed on users' Android devices or the configured emulators to capture the invocations of specified Android APIs when the target apps run in realtime, where the logged contents are elaborated to facilitate the log matching.
The logs are then sent to \textit{Matching Module} for the following analysis.

\noindent\textbf{Matching Module.} The module is run offline to reveal contextual information of app behaviors by the logs received from \textit{Logging Module}.
It obtains the APK file from app markets or the Internet according to the logs.
Then it achieves the log matching as follows.
It preprocesses the app code and the logs to output supergraphs and log segments respectively.
The supergraph combines CFGs and the callgraph (\ie, CG) of the app. 
Next, it positions each log point matched with a log record on the corresponding supergraph individually and then explores a path to connect the log points.
\section{Collecting \& Filtering of Logs}
\label{sec:log&app_analysis}

\begin{figure}[t]
        \centering

        \includegraphics[width=3.1in]{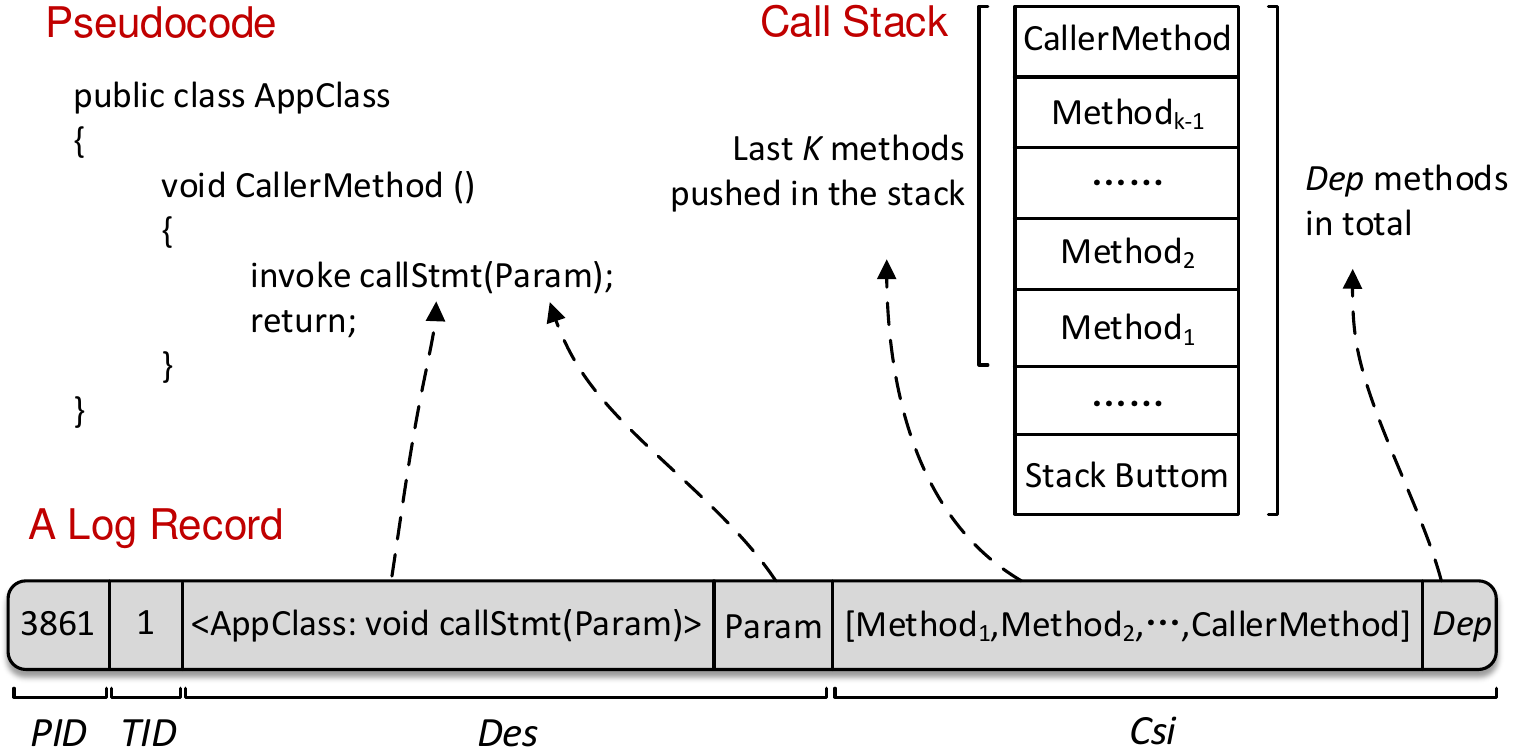}
        \caption{An example of the composition of a log record.}
        \label{fig:stack}
\end{figure}

This section first depicts the information that profiles app behaviors and needs to be logged for achieving the log matching.
There may be a large number of information unrelated to the target apps being logged in practice, so the section then describes how to remove the noise from log data, which ensures the precision and efficiency of the contextual reveal.

\subsection{Logged Information} \label{logginginformation}
We collect log information from the OS to prepare for the log matching.
Specifically, a time-stamped sequence of log records is represented as $L$.
Since $L$ is temporally ordered, we use the sequence without timestamps to model $L$.
A log record is a tuple $\langle$\textit{PID}, \textit{TID}, \textit{Des}, \textit{Csi}$\rangle$.
\textit{PID} and \textit{TID} indicate the process and the thread printing the log record.
\textit{Des} contains the invoked API's descriptions including the signature of the logged API, the used arguments, the reture value, \etc
\textit{Csi} is the information extracted from the runtime call stack when we log the API.
\Cref{fig:stack} depicts an example of the composition of a log record, and the details are shown as follows.

\begin{itemize}
\item \textit{PID} is used to bind an app to the log records that the app produces, and \textit{TID} is used to distinguish the log records output by different threads.

\item \textit{Des} is used to obtain the invoked API, to parse the API when it involves to some mechanisms, \eg, reflection and ICC, and to distinguish the same API calls with different arguments or return values.
Note that the selection of the logged information depends on analysts' requirements, which is discussed in \Cref{sec:evaluation} and \Cref{sec:limitations}.

\item \textit{Csi} is the crucial information to position the log point matched with the log record in the CFG of the target app.
As shown in \Cref{fig:stack}, it includes the last \textit{K} methods pushed in the call stack~(\textit{Las}) and the depth of the stack (\textit{Dep}).
Specifically, \textit{Las} indicates the method call sequence for finding the method that contains the log point matched with the log record.
\textit{Dep} presents the number of invoked methods in the stack.
With the elements, the log point can be located in a method.
\end{itemize}

We decide the value of \textit{K} to balance the effectiveness of the log matching and the imposed runtime overhead.
On the one hand, the methods that are close to the stack bottom do not need to be logged because they are related to the OS for initialization tasks (\eg, \texttt{com.android.internal.os.ZygoteInit()}) rather than the target app code.
In other words, there is no node matched with the methods in the CG built on the app bytecode.
Furthermore, considerable overhead is imposed when collecting overmuch information from the stack for each log record.
On the other hand, a log point may not be precisely positioned if the sequence from the entry point to the method containing the log point is not recorded entirely.
The trade-off is discussed in \Cref{sec:evaluation} to decide the value of \textit{K}.

\subsection{Log Filtering} \label{logfilter}
To ensure that the log matching is not affected by noise, we remove the log records produced by the OS, nontarget apps, the implementation of Android framework APIs, \etc, from \textit{L}.
Since maliciousness of target apps is usually hidden in the code of host apps and embedded third-party libraries~\cite{li2017understanding}, we focus on revealing contextual information from them rather than the aforementioned noise.
As mentioned in \Cref{logginginformation}, \textit{PID} is used to remove the log records produced by the processes whose process IDs are different from the target app.
However, this approach cannot be used to identify the log records produced by the implementation of Android framework APIs used in the target app.
For example, the implementation of \texttt{onCreate()} requires to call the API named \texttt{<java.lang.Class: java.lang.Object newInstance()>}.
The log record of the API call is not our concern but has the same \textit{PID} as the target app.
We adopt a solution that removes the log records based on class names of the methods that contain the corresponding log points (\eg, \texttt{AppClass} is the class name of \texttt{CallerMethod()} in \Cref{fig:stack}).
The class names indicate if the APIs are invoked by the code of Android framework libraries.
For instance, \texttt{android.app.$*$} and \texttt{java.security.*} are the prefixes of class names of the Android framework libraries.

\begin{table}[t]
\centering
\caption{The top-3 frequently-occurring APIs in the log files of 8 apps.}
\label{tab:space_overhead}
\resizebox{\linewidth}{!}{
\begin{tabular}{|l|l|l|l|}
\hline
\textbf{App Name} & \textbf{First-ranked API} & \textbf{Second-ranked API} & \textbf{Third-ranked API} \\
\hline
Chrome & \texttt{dispatchMessage}/56\% & \texttt{getCountry}/26.3\% & \texttt{getText}/7.8\% \\
\hline
TencentMap & \texttt{dispatchMessage}/56.2\% & \texttt{newInstance}/15.6\% & \texttt{invoke}/6.8\% \\
\hline
Note & \texttt{dispatchMessage}/56.6\% & \texttt{write}/21.3\% & \texttt{getText}/8.5\% \\
\hline
SogouInput & \texttt{dispatchMessage}/68.9\%  & \texttt{newInstance}/9.6\% & \texttt{getText}/7.1\% \\
\hline
YoudaoDict & \texttt{dispatchMessage}/49.4\%  & \texttt{invoke}/13\% & \texttt{run}/11.5\% \\
\hline
Androidesk & \texttt{dispatchMessage}/51.2\%  & \texttt{getText}/10.5\% & \texttt{newInstance}/10.4\% \\
\hline
RipTide GP & \texttt{dispatchMessage}/88.6\%  & \texttt{getText}/1.5\% & \texttt{getCountry}/1.2\% \\
\hline
HungryShark3 & \texttt{dispatchMessage}/84.1\%  & \texttt{newInstance}/3.8\% & \texttt{invoke}/2.8\% \\
\hline
\end{tabular}
}
\end{table}

We experiment to validate the necessity of the log filtering with the configuration in \Cref{sec:evaluation}-A.
\Cref{tab:space_overhead} lists the top-3 frequently-occurring logged APIs and their proportions in the log files generated by 8 apps without the log filtering.
We find that more than 50\% of logged APIs are \texttt{dispatchMessage()}.
After further investigation, we know that the API is invoked by the code of the Android framework library named \texttt{android.os.Looper} at the system level.
Similarly, \texttt{newInstance()} and \texttt{getText()} are commonly invoked by the code of \texttt{android.view} for UI operations in the experiment.
We need to filter the log records because they are a large part of the output logs, but are not our concern for the contextual reveal in apps.

Note that our main contribution is achieving the precise and efficient log matching for revealing contextual information of app behaviors, instead of designing novel logging and filtering techniques.
Therefore, we adopt the standard logging scheme as mentioned before.
The experimental results in \Cref{sec:evaluation} demonstrate that it is sufficient for our work to achieve the log matching and the contextual reveal by the adopted scheme.

\section{Log Matching} \label{sec:divide-and-conquer}

\begin{algorithm}[t]
  \caption{The implementation of our strategy.}
  \label{alg:matchingPath}
  \footnotesize
   \begin{algorithmic}[1]
        \Procedure{logMatching}{\textit{L}, CFGs, CG}
        \State{// Preprocessing}
        \State{Set$_{<tid,List_<ls, m>>}$ = logSplit(\textit{L})}
        \State{Set$_{<tid,List_{<ls, sg>>}}$ = genGraph(CFGs,CG,Set$_{<tid,List_<ls, m>>}$)}
        \State{// The divide and conquer strategy}
        \State{List$_{Ep}$ = $\varnothing$, Set$_{<tid, cp>}$ = $\varnothing$}
        \For{each $<tid, List_{<ls, sg>}>$ $\in$ Set$_{<tid, List_{<ls, sg>>}}$}
            \For{each $<$\textit{ls}, \textit{sg}$>$ $\in$ List$_{<ls, sg>}$}
                \State{List$_{Pos}$ = $\varnothing$}
                \For {each \textit{lr} $\in$ \textit{ls}}
                    \State{pos = logPointPosition(lr)}
                    \State{List$_{pos}$ = List$_{pos}$ $\cup$ pos}
                \EndFor
                \State{ep $\leftarrow$ pathExploration(\textit{sg}, List$_{pos}$)}
                \State{List$_{ep}$ = List$_{ep}$ $\cup$ ep}
            \EndFor
            \State{combinedPath = pathCombination(List$_{ep}$)}
            \State{Set$_{<tid, cp>}$ = Set$_{<tid, cp>}$ $\cup$ $<$$tid$, combinedPath$>$}
        \EndFor
        \EndProcedure
   \end{algorithmic}
\end{algorithm}

We depict how \oursystem reveals contextual information of app behaviors after obtaining the required logs in this section.
The output results are program paths that contain various details of app behaviors, on which security analysts are more likely to identify hidden maliciousness of the target apps.

\subsection{Preprocessing} \label{sec:processing}
In the following, we first explain how to split the log sequence \textit{L} received from \textit{Logging Module} and then introduce the graph structure used by \oursystem to identify the path matched with audit logs. 

\subsubsection{Log Splitting}
\oursystem splits \textit{L} into multiple segments based on \textit{TID} and the logged callback methods that are invoked by the Android framework \cite{DBLP:conf/pldi/ArztRFBBKTOM14}.
The log records of \textit{L} are produced by different threads, and meanwhile, log points matched with the log records are located in different CFGs.
Hence, the computational complexity is huge if performing the log matching on the intertwined logs directly.

To facilitate the log matching, we first split \textit{L} into multiple subsequences, each of which contains the log records with the same \textit{TID}.
In each subsequence, the produce of the log records between two successive logged callback methods stems from the invocation of the former method because the subsequence is temporally ordered.
Therefore, for each subsequence, we obtain a list of log segments each of which consists of a callback method and the log records between this method and the next logged callback method.

\begin{figure}[t]
    \centering 
    \includegraphics[width=0.9\linewidth]{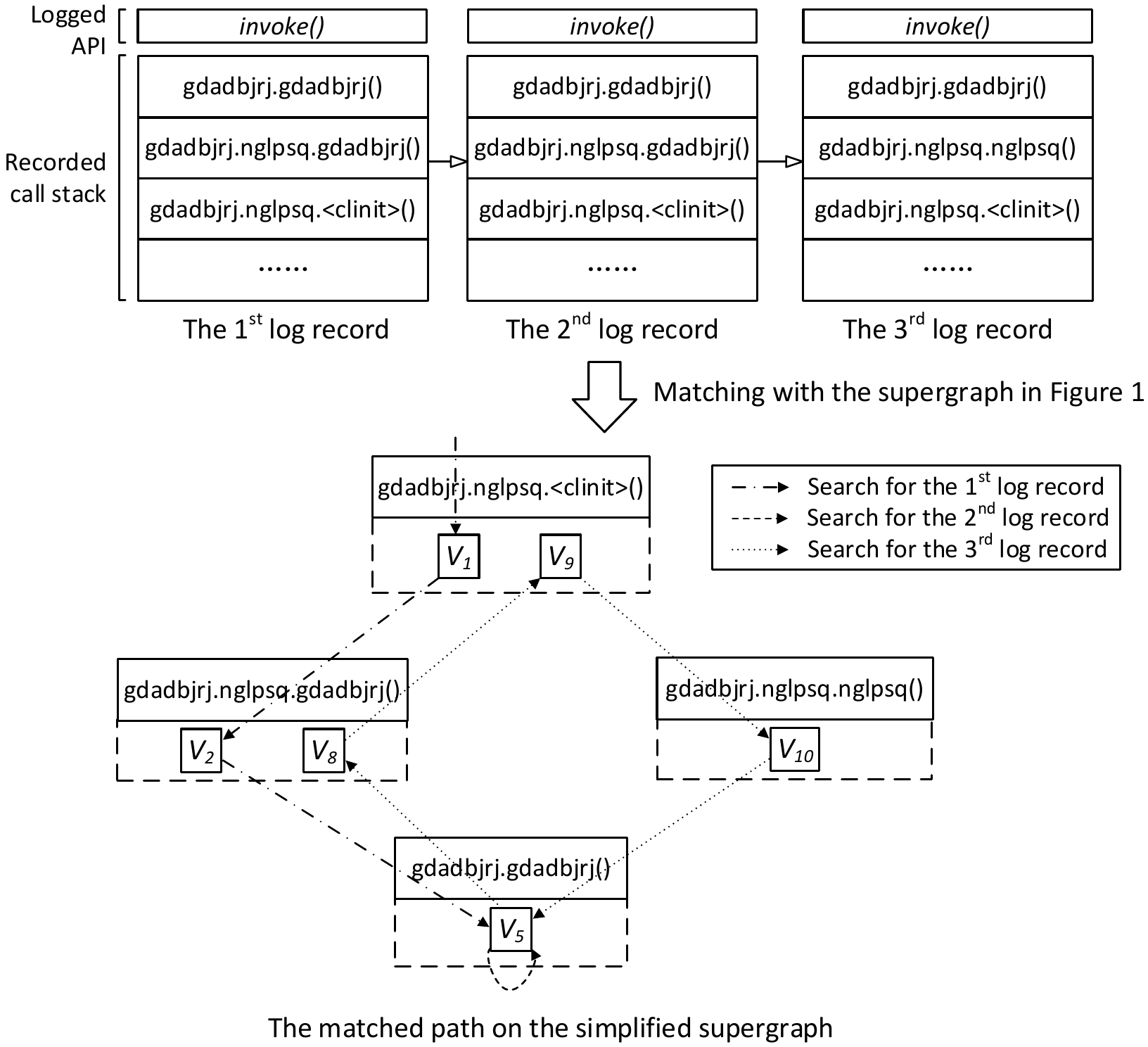}
    \caption{Matching a log segment with the graph in \Cref{fig:problemStatement} by our strategy.}
    \label{fig:divideandconquer}
\end{figure}

\subsubsection{Supergraph Construction}\label{subsubsection:graphbuilding}
We build supergraphs for each logged callback method based on the CFGs and the CG of the target app.
Unlike traditional console applications which have \texttt{main()} as entry points, there is no common entry point for Android apps.
They await orders from users (\eg, clicking a button), systems (\eg, in a low battery state), or other apps to launch a specific callback method.
Existing static analysis schemes \cite{DBLP:conf/pldi/ArztRFBBKTOM14,DBLP:conf/icse/OcteauLDJM15,DBLP:conf/icse/0029BBKTARBOM15} predefine the calling orders of the callback methods in a dummy-main method, which is regarded as the entry point of an app.
Nevertheless, we find that not all callback methods are collected in the dummy-main method.
For instance, a dummy-main method of an app named \textit{com.shuqi.paid.controller} built by FlowDroid misses a method named \texttt{com.shuqi.controller.Main.onStart()}. 
Furthermore, the sophisticated maintenance of the calling order of callback methods may be error-prone, which is illustrated in \Cref{sub:performance_on_test_suites}.

To solve the problem, \oursystem orderly extracts invoked callback methods from \textit{L} of the target app and builds a supergraph for each of them.
The root node of each supergraph is the head node of the CFG of a callback method, and a supergraph is built by combining the CFGs for all methods that are reachable from a callback method via the app's CG.
This scheme recognizes all the callback methods invoked at runtime and models a definite calling order for them, which complements the existing static analysis schemes.
Furthermore, a callback method corresponds to a supergraph, so we get the matching relation between each log segment and a supergraph based on the callback method of the log segment. 
Note that \oursystem aims to circumvent a part of the inherent defects of static analysis in building the precise graph structure based on the runtime information, instead of thoroughly solving the problem.

\subsection{Our Divide and Conquer Strategy}
\label{sub_logmatching}

We design a divide and conquer strategy to achieve the log matching by local search instead of using the backtracking search globally.
\Cref{alg:matchingPath} describes the implementation of our strategy for the log matching.
As explained in \Cref{sec:processing}, the algorithm splits the log sequence into multiple log segments based on \textit{TID} and the logged callback methods and then constructs supergraphs for the methods (lines 3-4).
By the preprocessing, the strategy divides the problem of matching a log sequence with the app code into multiple subproblems, each of which matches a log segment with a supergraph.
Each pair of matching relation is saved in List$_{<ls, sg>}$.
Next, the algorithm divides the subproblem into multiple subsubproblems, each of which positions the log point matched with a log record in the corresponding supergraph based on the logged call stack information (lines 10-12).
After the positions are decided, the algorithm explores the path corresponding to the runtime execution flow to connect the log points orderly in the supergraph (lines 13-14).
When the paths matched with each log segment are obtained, the algorithm combines them to generate the result (line 15).
Finally, the algorithm saves the mapping between \textit{TID} and the combined path in Set$_{<tid, cp>}$ (line 16).

\Cref{fig:divideandconquer} depicts the process of our strategy in matching a log segment with the graph snippet in \Cref{fig:problemStatement}.
We notice that the APIs of the log points matched with the first two log records are invoked along the same method call sequence, while the method call sequence for invoking the API of the log point matched with the third log record is different from them. 
Considering that the log records are ordered temporally in a program execution and the path extracted from the graph of the program is matched with the log sequence, the log points on the path are matched with the corresponding log records orderly.
Therefore, the APIs for the first two log records are continuously invoked in the method named \texttt{gdadbjrj.gdadbjrj()} at runtime, but the API for the third log record is invoked along another method call sequence.
As shown in \Cref{fig:divideandconquer}, our strategy explores the three log points along the three path segments orderly on the graph.

\begin{figure}[t]
    \centering 
    \includegraphics[width=0.9\linewidth]{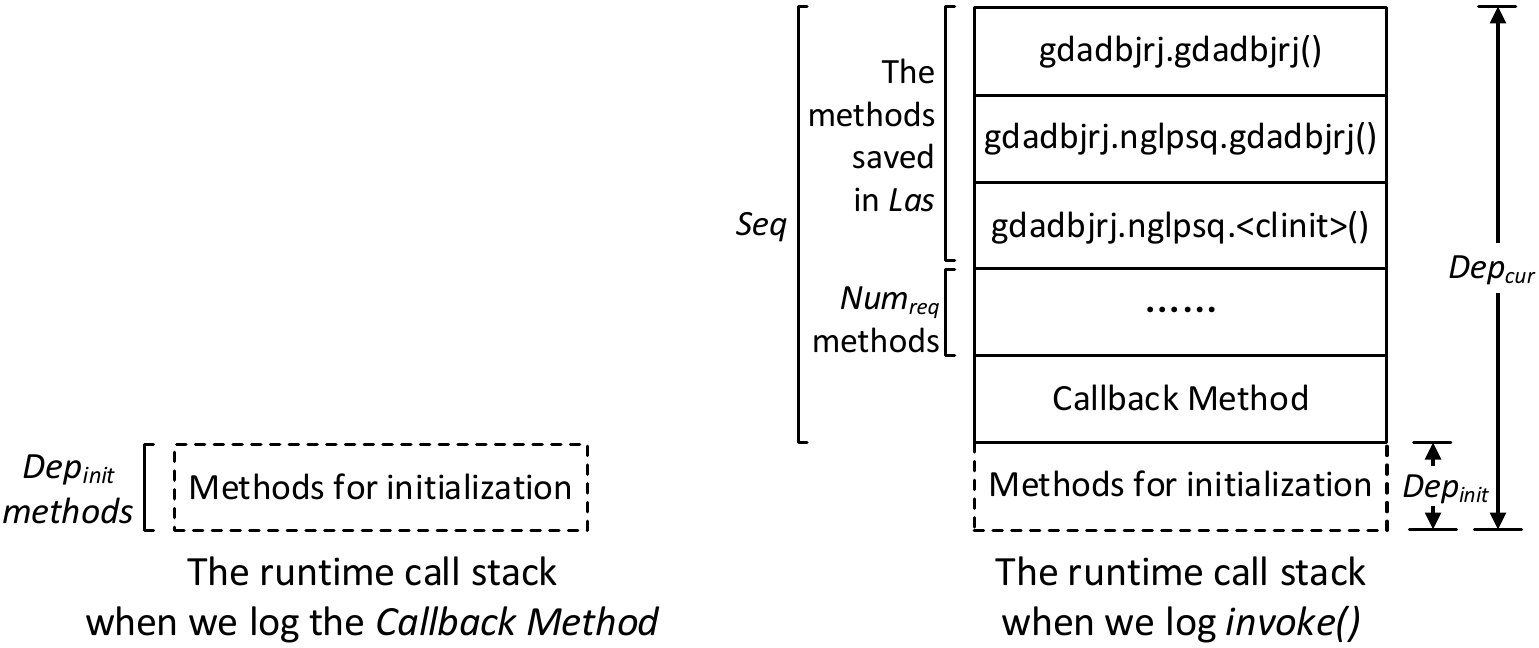}
    \caption{The details of the call stack when we log \texttt{Callback Method} and \texttt{invoke()} in \texttt{gdadbjrj.gdadbjrj()} respectively.}
    \label{fig:divideCase}
\end{figure}

\subsubsection{Positioning Log Points}
Our strategy finds the method that contains the log point matched with a log record in the supergraph based on the \textit{Csi} in the log record.
We describe the position of a log point by a tuple $\langle$\textit{Num$_{req}$},~\textit{Las}$\rangle$.
As depicted in \Cref{fig:divideCase}, \textit{Num$_{req}$} is the number of methods that are pushed in the runtime call stack after the callback method (\eg, \texttt{Callback Method}) but are not included in \textit{Las}, and \textit{Seq} is the method call sequence from the callback method to the method that contains the log point (\eg, \texttt{gdadbjrj.gdadbjrj()}) in the stack.
We calculate \textit{Num$_{req}$} by the following equation:
$$\textit{Num$_{req}$=}
\begin{cases}
\textit{Dep$_{cur}$}~-~\textit{Dep$_{init}$}~-~\textit{K}~-~1& \text{if \textit{Las} $\subsetneq$ \textit{Seq}}\\
0& \text{if \textit{Seq} $\subseteq$ \textit{Las}}
\end{cases}$$
, where \textit{Dep$_{cur}$} is the depth of the call stack when we log the API invoked at the log point (\eg, \texttt{invoke()} in \Cref{fig:divideCase}), \textit{Dep$_{init}$} is the depth of the call stack when we log the callback method, and \textit{K} is the length of \textit{Las}.

Based on this equation, if \textit{Seq} is a subset of \textit{Las}, our strategy positions the log point along the method call sequence in \textit{Seq} without \textit{Num$_{req}$}.
If \textit{Las} is a proper subset of \textit{Seq}, the strategy uses \textit{Num$_{req}$} to help position the log point.
Specifically, there may be multiple method call sequences between the callback method and the first method of \textit{Las} (\eg, \texttt{gdadbjrj.nglpsq.<clinit>()} in \Cref{fig:divideCase}) on the supergraph.
Hence, the strategy leverages \textit{Num$_{req}$} to indicate the number of methods that need to be traversed from the callback method before searching along \textit{Las}.

\begin{algorithm}[t]
  \caption{The implementation of the path exploration.}
  \label{alg:pathExploration}
  \footnotesize
   \begin{algorithmic}[1]
        \State{Stack$_{m}$ = $\varnothing$}
        \Procedure{pathExploration}{\textit{N}}
        \If{\textit{N} is a head node of method \textit{m}}
            Stack$_{m}$.push(\textit{m})
        \EndIf
        \If{Stack$_{m}$.length() \textless\ \textit{Num$_{req}$} + 2}
            \For{each \textit{n} $\in$ \textit{N}.successors}
                \Call{pathExploration}{\textit{n}}
            \EndFor
        \EndIf
        \If{Stack$_{m}$.length() $\geq$\ \textit{Num$_{req}$} + 2}
            \State{\textit{d} = Stack$_{m}$.length() - (\textit{Num$_{req}$} + 1})
            \If{Stack$_{m}$.sub(\textit{Num$_{req}$} + 1,Stack$_{m}$.length()) == \textit{Las}.sub(0,\textit{d})}
                \If{\textit{d} == \textit{Las}.length} Find a log point in \textit{N} \EndIf
                \For{each \textit{n} $\in$ \textit{N}.successors}
                    \Call{pathExploration}{\textit{n}}
                \EndFor
            \Else \If{\textit{N} is a head node} Stack$_{m}$.pop()
                    \Else
                    \If{\textit{N} is a tail node} Stack$_{m}$.pop()
                    \EndIf
                    \For{each \textit{n} $\in$ \textit{N}.successors} \Call{pathExploration}{\textit{n}}
            \EndFor
                  \EndIf
            \EndIf
        \EndIf
        \EndProcedure
   \end{algorithmic}
\end{algorithm}

\subsubsection{Exploring the Path}
Our strategy explores the path corresponding to the runtime execution flow by orderly connecting the log points matched with a log segment.
Overall, the strategy addresses the log matching during the path exploration according to the consistency between the call stack recorded when running the target program and the method call sequence generated when traversing the supergraph.
Specifically, we need to solve two problems as follows.

The first problem is how to explore the path with as few backtracks as possible.
As explained in \Cref{problemstatement}, straightforward searches may cause high computation complexity.
Therefore, we design \Cref{alg:pathExploration}, in which we aim to ensure the consistency of three pairs of elements during the log matching: 1) the length of Stack$_{m}$ and the length of \textit{Seq} shown in \Cref{fig:divideCase},
2) the API invoked by the log {}point and the API in the log record, and 3) the last \textit{K} methods pushed in Stack$_{m}$ and \textit{Las} in the log record.

The algorithm saves the methods being visited on the explored path in Stack$_{m}$ one by one.
If the length of Stack$_{m}$ is less than (\textit{Num$_{req}$} + 2), it performs the pure backtracking search to increase the length (lines 4-5).
When the length of Stack$_{m}$ is greater than or equal to (\textit{Num$_{req}$} + 2), it searches with keeping that the method sequence in Stack$_{m}$ is consistent with \textit{Las} (lines 6-8).
After the method that contains a log point is visited, it searches the method's CFG to find the matched log point (lines 8-10).
If the method sequence in Stack$_{m}$ is inconsistent with \textit{Las}, it stops exploring the method at the top of Stack$_{m}$ (line 12) or searches back to the visited methods along the return edges of the supergraph (lines 14-15).
The path exploration finishes when all log points are connected.

Our strategy addresses the log matching for recursive calls.
With the aforementioned consistency, the strategy differentiates the log points invoking the APIs at different recursion depths.
Even if the APIs are the same or some methods in the runtime call stack are not logged due to the insufficient length of \textit{Las}, the strategy searches the method recursively or exits from a recursion based on the difference of the length between Stack$_{m}$ and \textit{Seq}.

\begin{figure}[t]
    \centering 
    \includegraphics[width=0.8\linewidth]{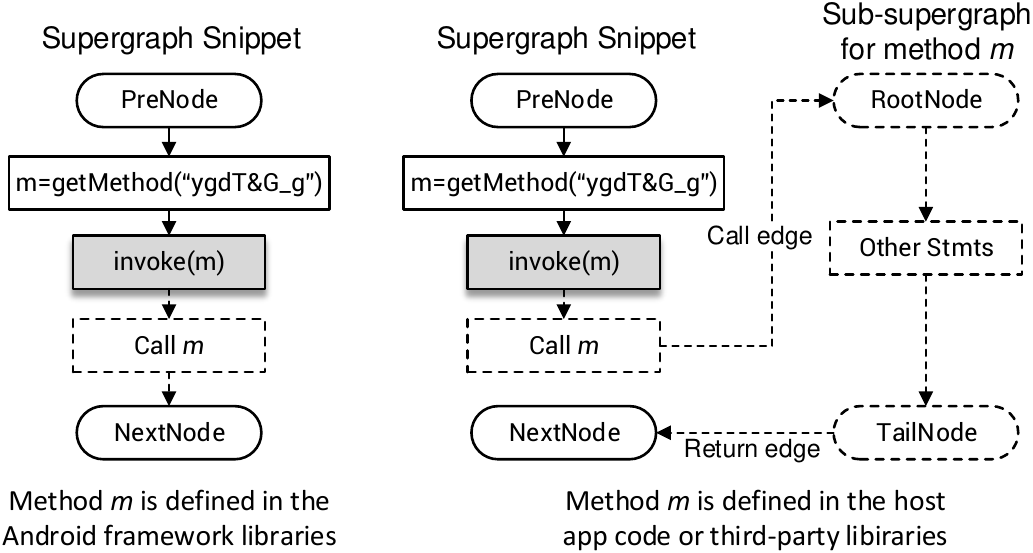}
    \caption{Two cases of updating the supergraph snippet for \texttt{invoke()}.}
    \label{fig:VCFG_Update}
\end{figure}

The second problem is how to decide the successors of the nodes, where the represented APIs involve some Android mechanisms (\eg, ICC, reflection) that make callee methods be determined dynamically.
Supergraphs built by static analysis are imprecise because of unresolved reflective calls~\cite{li2016droidra}, imprecise ICC links \cite{DBLP:conf/icse/OcteauLDJM15}, \etc
To circumvent a part of the inherent defects of static analysis, our strategy updates the successors of the nodes according to the used arguments in log records.
It supports handling many cases for reflection, ICC, \etc.
Due to the page limit, we show a solution for \texttt{invoke()} as follows.
Our strategy obtains the signature of the invoked method from the recorded arguments and then updates the supergraph based on the two types of methods.
\begin{itemize}
\item The method defined in the Android framework libraries, \eg, \texttt{sendTextMessage()}.
As shown in the left part of \Cref{fig:VCFG_Update}, our strategy inserts a new node, where the statement calls the method \textit{m} explicitly, in the supergraph.

\item The method defined in the host app code or third-party libraries, \eg, \texttt{gdadbjrj.gdadbjrj()} in \Cref{fig:divideandconquer}.
Our strategy not only inserts a new node but also embeds a sub-supergraph in the original supergraph where the root node is in the head node of the method's CFG.
\end{itemize}

\subsubsection{Combining Explored Paths}
We obtain the final results matched with \textit{L} by combining the explored paths according to the \textit{TID} and the logging sequence of callback methods.
We first connect the nodes of the paths according to the execution orders where the same nodes are merged, if there are multiple paths matched with a log segment.
Next, for the log segments with the same \textit{TID}, we orderly connects the last node of the path matched with the previous log segment to the first node of the path matched with the next log segment.
\section{Experimental Evaluation} 
\label{sec:evaluation}

To evaluate the effectiveness of \oursystem, we seek to answer the following questions:

\begin{itemize}
\item \textbf{RQ1:} How is the performance overhead and space consumption imposed by \textit{Logging Module} of \oursystem?
\item \textbf{RQ2:} Whether our log matching can handle various scenarios of Android apps in micro-benchmarks?
\item \textbf{RQ3:} How precise and efficient is our log matching on real-world apps?
Whether contextual information is correctly revealed via \oursystem for real-world apps?
Whether the revealed details of AppAngio complement the reports of existing techniques?
\item \textbf{RQ4:} Whether AppAngio helps analysts disclose maliciousness of app behaviors?
How to make experts analyze the matched paths and retrieve information efficiently?  

\end{itemize}

\subsection{Experimental Setup}
\subsubsection{Implementation} We implement a prototype of \oursystem.
\textit{Logging Module} can be deployed on real devices or emulators with different Android versions, and here we choose to modify the source code of Android 5.0.1 to achieve the logging mechanism and flash its system image into the device of Nexus 4.
Moreover, \textit{Matching Module} is developed on Soot~\cite{lam2011soot} and extends some off-the-shelf interfaces~\cite{DBLP:conf/pldi/ArztRFBBKTOM14,li2016droidra,DBLP:conf/icse/0029BBKTARBOM15} to generate the supergraphs and rebuild the call relations for reflection, ICC, \etc
The module is deployed on a server with Intel Broadwell E5-2660V4 2.0GHz CPU, 128G memory and Ubuntu 16.04 LTS (64 bit).

\subsubsection{Datasets} We randomly select 3000 market apps from AndroZoo \cite{allix2016androzoo} and also collect 3000 malware samples, including 1000 samples from MalGenome project \cite{DBLP:conf/sp/ZhouJ12} and 2000 samples from VirusShare~\cite{virusshare}. 
Then we also choose the open-source test suites of DroidBench \cite{DBLP:conf/pldi/ArztRFBBKTOM14} and ICC-Bench \cite{DBLP:conf/ccs/WeiROR14} as our micro-benchmarks.

\subsubsection{Selection of Logged APIs} \label{logselection}
\oursystem supports the flexible selection of logged APIs, which means that security analysts can set the APIs in their released Android OSes based on the requirements of security analysts.
Note that the analysts do not need to make major modifications to the OS, because AppAngio only logs the API calls rather than low-level system operations.
In this experiment, we log 166 APIs about privacy leaks by reviewing previous literature~\cite{DBLP:conf/pldi/ArztRFBBKTOM14,DBLP:conf/icse/0029BBKTARBOM15,gordon2015information}, because the analysis of privacy-breaching behaviors is one of the most important tasks for the Android security~\cite{DBLP:conf/pldi/ArztRFBBKTOM14}.
The APIs are under \texttt{framework/}, \texttt{libcore/} and \texttt{art/} directories of the Android OS source code and in the following three types.

\noindent\textbf{Event Handlers.} The event handlers are a series of callback methods related to the state transitions of Android lifecycles, GUI operations and system events \cite{DBLP:conf/ccs/YangYZGNW13}.
An event handler contains the semantic of behavioral activation.

\noindent\textbf{Privacy-related Operations.} These are the operations of privacy leaks.
We choose the frequently-used operations from sources or sinks specified in SuSi \cite{DBLP:conf/ndss/RasthoferAB14}.
The logged operations are divided into the following categories: account, bluetooth, device information, database, file, network, SMS, \etc

\noindent\textbf{ICC and Reflection-based Operations.} These logged operations are used to update supergraphs.
Specifically, we record the origin, the target and the invoked API for each ICC link.
We also record the reflection-based APIs and their used arguments.
The arguments indicate dynamically loaded classes, invoked methods, \etc

\oursystem can log native APIs invoked via JNI by adding log printing statements at the C/C++ level in the files under \texttt{libnativehelper/} directory of our Android OS source code.
For example, it logs the APIs named \texttt{GetMethodID()}, \texttt{CallVoidMethod()} and \texttt{CallObjectMethod()}, \etc, in \texttt{jni.h} file to obtain the signatures and the arguments of the invoked Java APIs.
Similarly, it can log different native APIs invoked via JNI even if native code is obfuscated.
Since the Soot framework is not capable of processing native code, neither as source code nor as binaries~\cite{arzt2017static}, we do not experiment to reveal contextual information of app behaviors in native code temporarily, which is discussed in \Cref{sec:limitations}.

\begin{figure}[t]
    \centering 
    \includegraphics[width=0.85\linewidth]{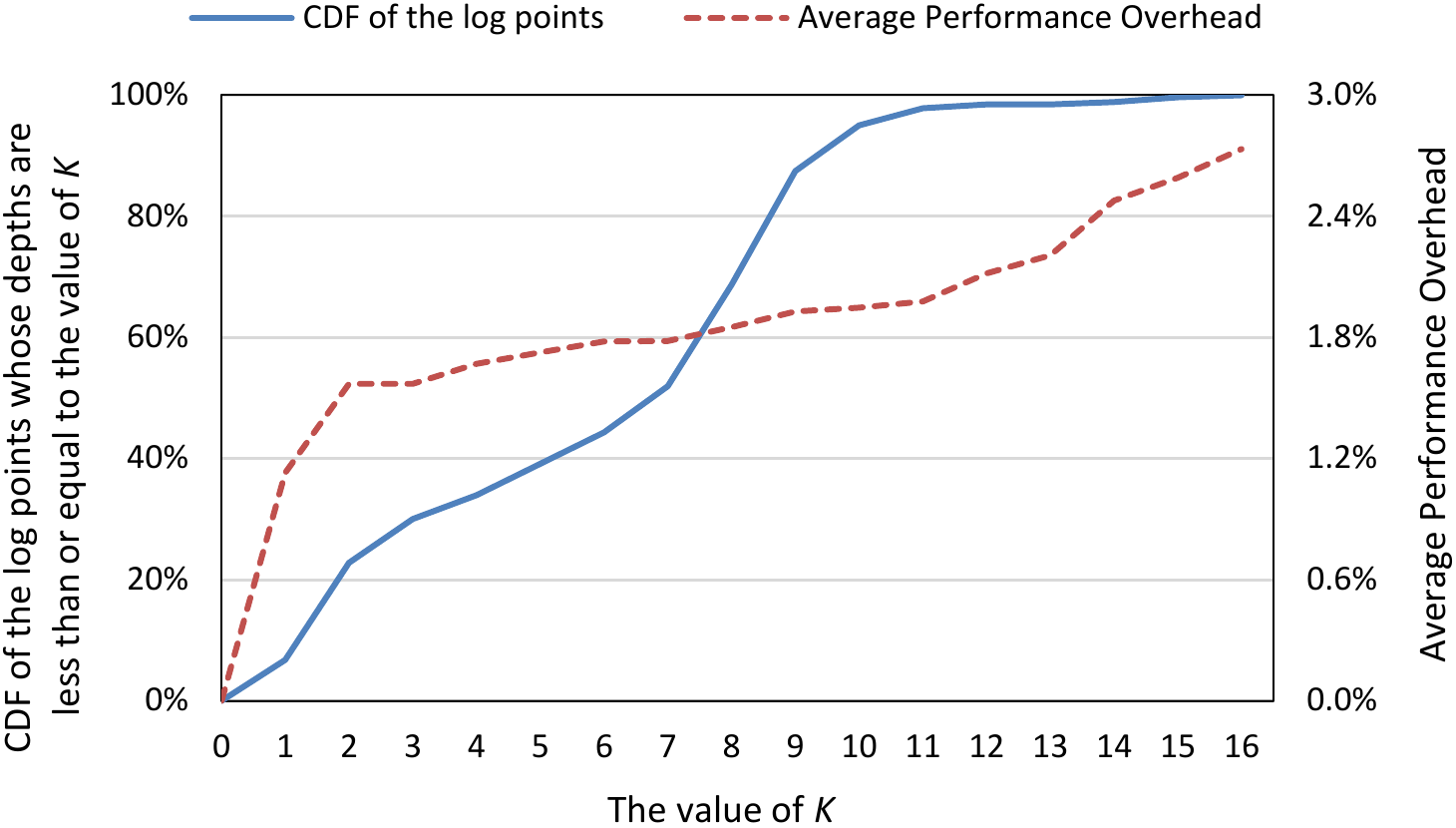}
    \caption{The statistics under different values of \textit{K}.}
    \label{fig:logPerformance}
\end{figure}

\subsubsection{Deciding the value of \textit{K}} \label{decideK}
Since the value of \textit{K} is related to practical factors (\eg, the selection of logged APIs, the performance of the device), we perform empirical experiments as follows.
To balance the effectiveness of the log matching and the runtime performance of the Android device as explained in \Cref{logginginformation}, we aim to find the maximum ratio of the coverage of the log points whose depths are less than or equal to the value of \textit{K} and the performance overhead of the device under different values of \textit{K}.
Here, the depth is the longest length of the method call sequence from the entry point to the method that includes a log point on an app's callgraph.
When the depth is less than or equal to the value of \textit{K}, the method call sequence is completely saved in a log record.
Moreover, when the value of \textit{K} increases, more call stack information is logged, so the performance overhead of the device also increases.

We obtain the aforementioned coverage and performance overhead from the apps and the Android device respectively.
On the one hand, for each app in our datasets, we first collect the depths of the log points of the selected APIs in the corresponding callgraph.
We then calculate the cumulative distribution function (CDF) of the APIs that can be positioned under different values of~\textit{K}.
On the other hand, we use a benchmark named AnTuTu to measure the performance overhead of the smartphone under different values of \textit{K}, where the measured smartphone is not installed any third-party app. 
AnTuTu measures the device automatically and comprehensively, which reduces our burden on designing test methods and collecting related data separately.
The adoption of AnTuTu in many literature~\cite{yan2012droidscope,yuan2017droidforensics,enck2014taintdroid} also demonstrates its reliability.
Therefore, the score reported by AnTuTu is our reference to assess the performance of the device.
Under each \textit{K}, we run AnTuTu five times on the device.
After each run, AnTuTu output a total score that is the sum of the scores of four detailed test scenarios (\ie, CPU, GPU, Memory, UX), where the higher scores present better device performance.
We calculate the average of the total scores for each \textit{K} and obtain the overhead by comparing it with the average score of the device without the runtime logging.

\begin{figure}[t]
    \begin{minipage}{4.25cm}
        \includegraphics[width=4.25cm]{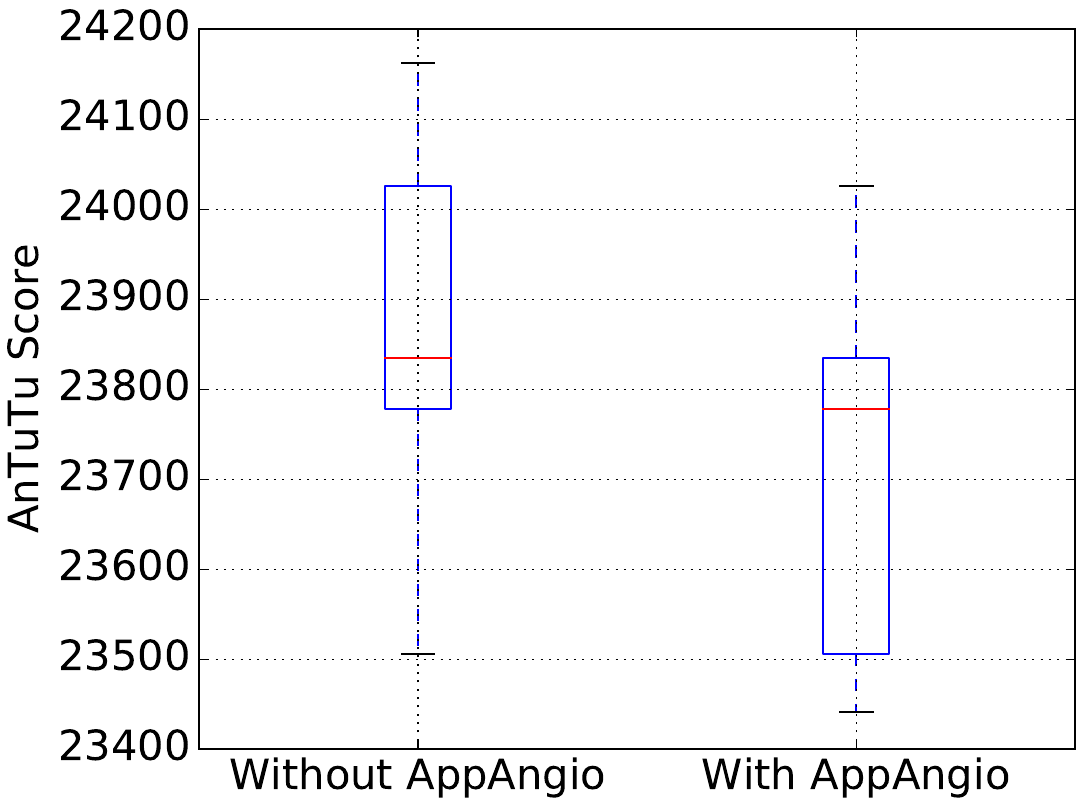}
        \centerline{\footnotesize{(a) The scores for CPU test}}
    \end{minipage}
    \begin{minipage}{4.25cm}
        \includegraphics[width=4.25cm]{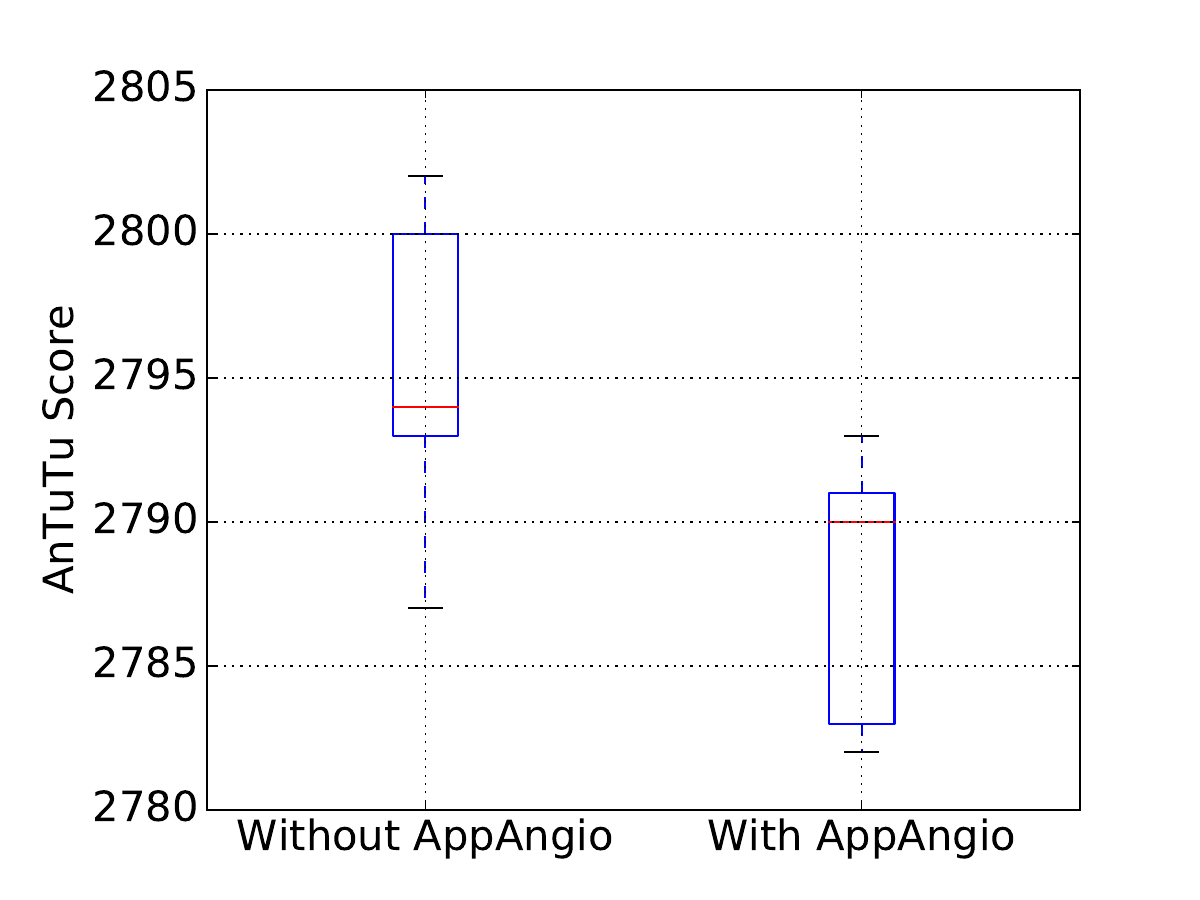}
        \centerline{\footnotesize{(b) The scores for GPU test}}
    \end{minipage}

    \begin{minipage}{4.25cm}
        \includegraphics[width=4.25cm]{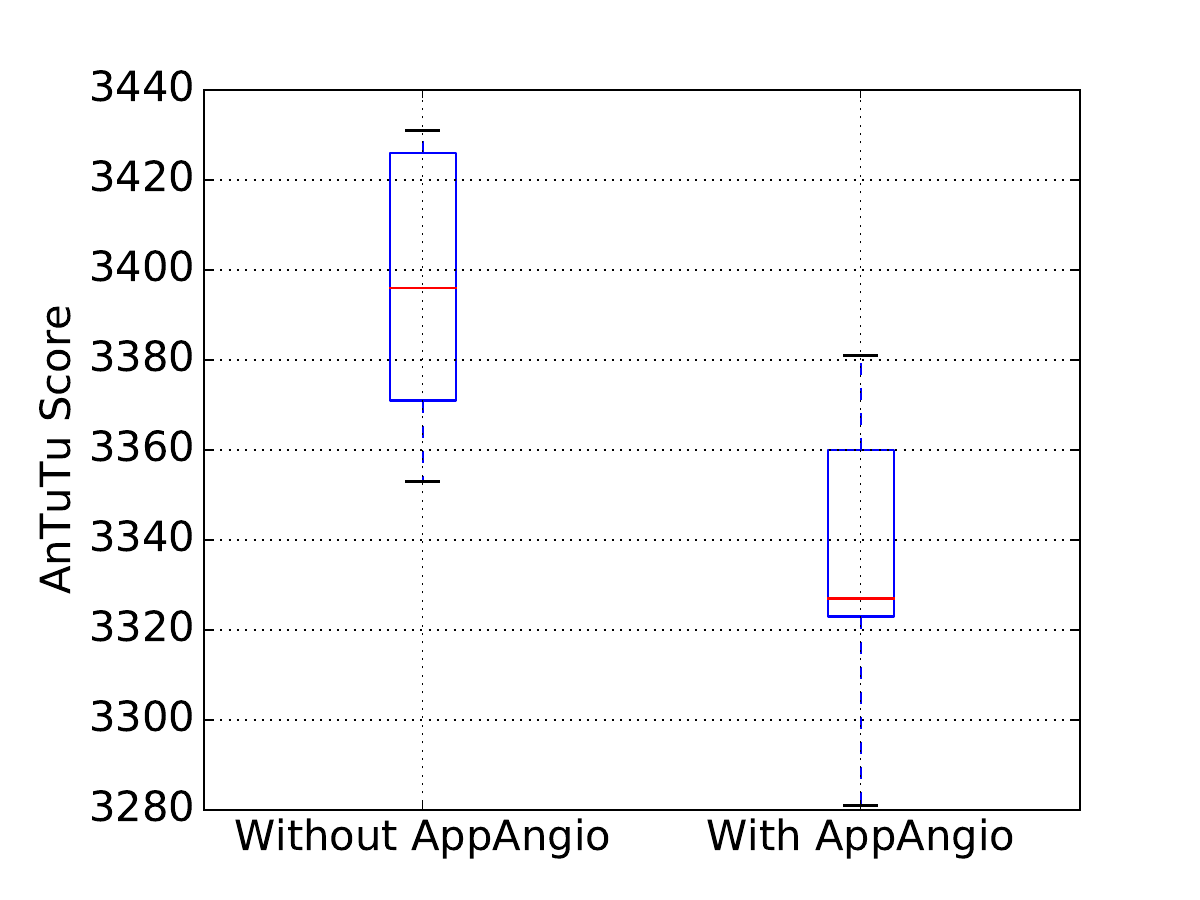}
        \centerline{\footnotesize{(c) The scores for Memory test}}
    \end{minipage}
    \begin{minipage}{4.25cm}
        \includegraphics[width=4.25cm]{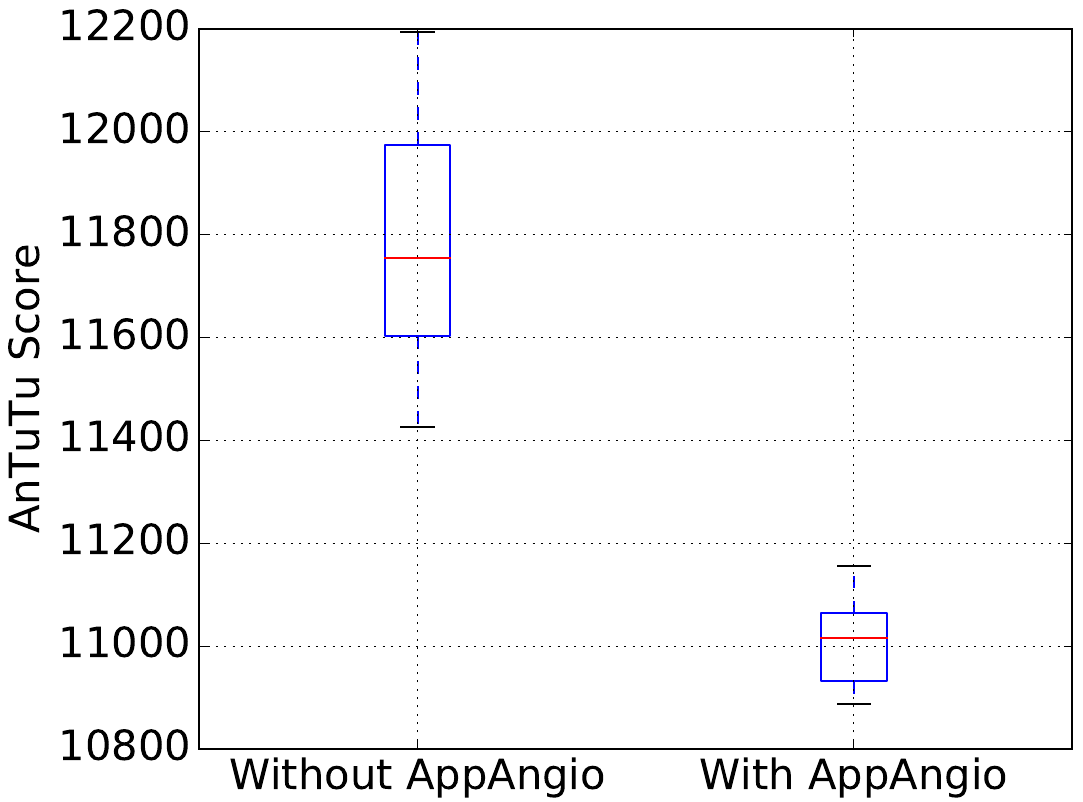}
        \centerline{\footnotesize{(d) The scores for UX test}}
    \end{minipage}
    \caption{The scores of four detailed test scenarios of AnTuTu.}
    \label{fig:systemOverhead}
\end{figure}

Based on the collected data, we fix \textit{K} as 11 in our evaluation.
\Cref{fig:logPerformance} depicts the statistics of the CDF and the average performance overhead under different values of \textit{K}, where the device is not equipped with \textit{Logging Module} when \textit{K} is 0.
Specifically, the depths of most log points (97.88\%) are less than or equal to 11, and the average performance overhead gradually increases (from 1.13\% to 2.73\%) when the value of \textit{K} increases from 1 to 16.
When \textit{K} is 11, the ratio of the aforementioned two indicators is the biggest.

\subsection{RQ1: Performance Overhead and Space Consumption} 
\label{sub:balance_between_logging_efficiency_and_judging_}

\subsubsection{Performance Overhead}
We analyze the performance of the device based on the experimental results in \Cref{decideK}.
The experiments show that our logging module that aims to collect API-level logs incurs negligible performance overhead (1.98\% on average).

\Cref{fig:systemOverhead} depicts the detailed results of the device on the four test scenarios of AnTuTu with and without \oursystem respectively.
Specifically, for the tests of CPU, GPU and Memory, the changes of the median value of scores when \oursystem is deployed on the device are negligible (less than 70), while the change for the UX test is 739.
After further analysis, we find that the change depends on the selection of logged APIs.
The UX test intensively performs the operations of I/O, data processing, \etc, which makes the involved APIs be logged frequently.
In daily use, we do not feel the affect of \oursystem on the performance of the device.

\begin{figure}[t]
    \centering 
    \includegraphics[width=0.85\linewidth]{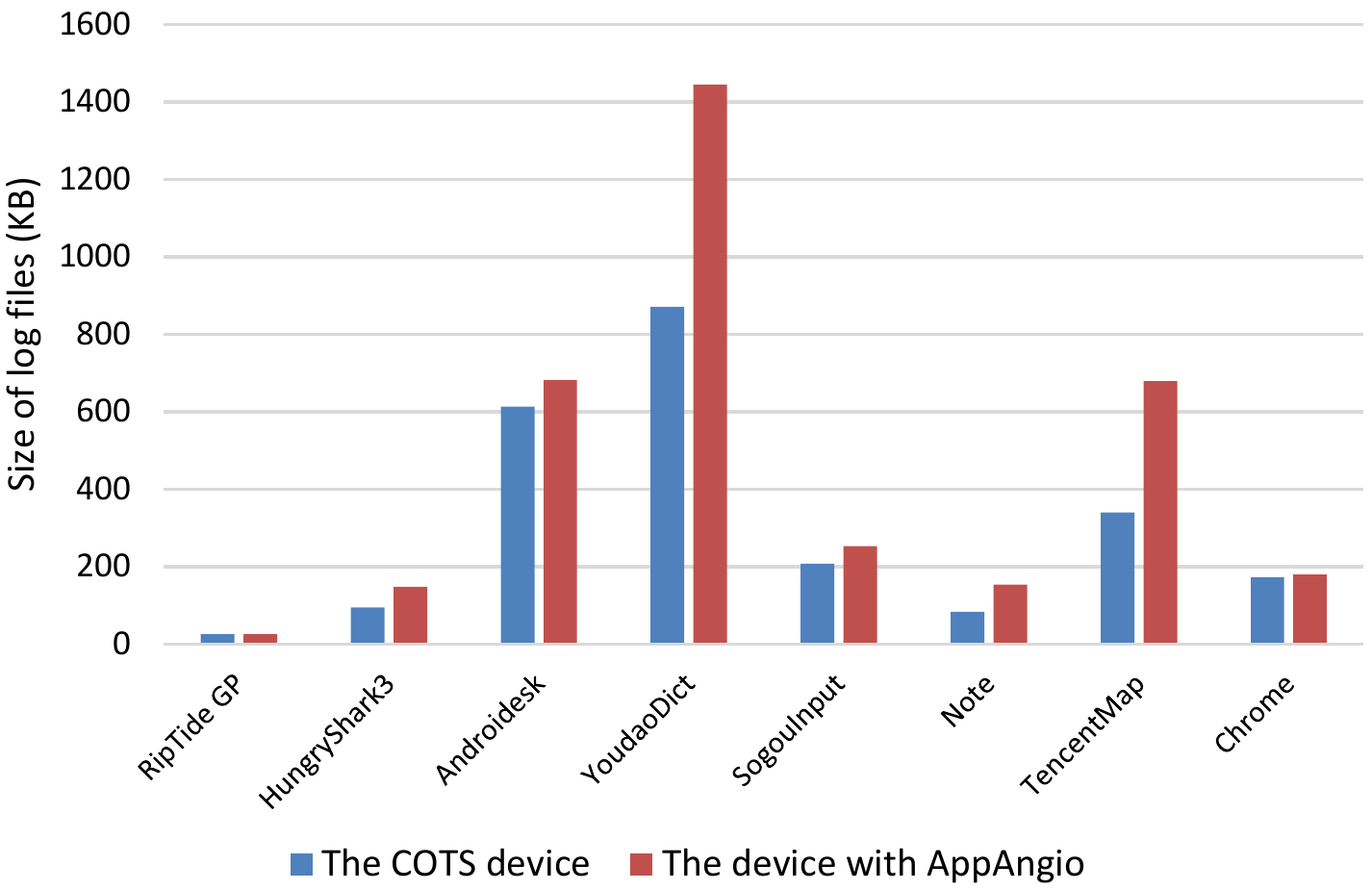}
    \caption{The statistics of the log file sizes for the 8 apps.} 
    \label{fig:logtest}
\end{figure}

\subsubsection{Space Consumption} To analyze the space consumption of our logging scheme, we randomly install 8 apps including Chrome web-browser, Tencent Map, and some other utility and game apps on the device and run them separately.
We manually operate each app up to 20 minutes to cover its functionalities as many as possible and obtain the log file via the command “adb logcat”.
To exhibit the space consumption of our logging scheme intuitively, we make a comparison with the log file sizes in the COTS device and the device with AppAngio.
To get the log files of the COTS device, we first obtain the log files from the device with AppAngio and then remove all the log records generated by \textit{Logging Module} of AppAngio.

The comparative results are shown in \Cref{fig:logtest}.
We notice that the space overhead of AppAngio, which increases the log file size by 144 KB on average, is related to various factors, \eg, the functionalities of testing apps, the embedded libraries, the logged APIs.
Note that the log files from the device with AppAngio are generated when the apps are operated intensively, so the log file sizes in \Cref{fig:logtest} are close to the worst-case results.
Moreover, to make a comprehensive evaluation for AppAngio in experiments, we log 166 APIs, which is more than the number of logged APIs reported by existing tools (\eg, DroidForensics \cite{yuan2017droidforensics} logs 21 Android APIs).
When fewer APIs are selected to log, the space overhead is reduced.

According to the aforementioned analysis, security analysts can choose to customize AppAngio in practice as follows.
For example, AppAngio can be designed to gather and upload log data periodically, and meanwhile, analysts can set the upper limit of the log file size and allow to overwrite the file when the file size exceeds the limit.
AppAngio can also maintain a whitelist configured by analysts to avoid logging for some trusted apps (\eg, \textit{YoudaoDict} or \textit{TencentMap} developed by the well-known enterprises).
Moreover, AppAngio can output the log records with a more compact format.

\subsection{RQ2: Effectiveness on Micro-benchmarks} 
\label{sub:performance_on_test_suites}
We regard the cases in the micro-benchmarks as the ground truth to examine if AppAngio can handle a variety of scenarios of Android apps.
DroidBench and ICC-Bench are the widely-adopted test suites to measure the effectiveness of Android analysis tools.
The cases include various programming schemes of Android apps involving different Android-specific challenges (\eg, modeling lifecycle, asynchronous callbacks, ICC).
Furthermore, the amount of code in each app is small (\ie, 56 LOC on average for each app), so it is suitable to analyze the code of each app detailedly.
We exclude 6 apps from DroidBench because they are designed for testing in Android emulators and inter-app communication, which are not considered in \oursystem temporarily.
Another excluded app in ICC-Bench cannot normally run due to the illegal integer assignment.
Therefore, we choose 113 cases in DroidBench and 19 cases in ICC-Bench for the evaluation.

\begin{table}[t]
    \caption{The representative detection results on micro-benchmarks
    ($\true$ = True Positive, $\truenegative$ = True Negative, $\falsepositive$ = False Positive, $\falsenegative$ = False Negative, DB = DroidBench, IB = ICC-Bench).}
    \label{tab:comparision_flowdroid_iccta}
    \centering

        \resizebox{0.485\textwidth}{!}{
    \begin{tabular}{|l|c|c|c|}
    \hline
    \textbf{AppName} & \textbf{FlowDroid + IccTA} & \tabincell{c}{\textbf{FlowDroid + AppAngio} / \\ \textbf{FlowDroid + IccTA + AppAngio}} & \textbf{AppAudit} \\
    \hline

    \multicolumn{4}{|c|}{\textbf{General Java}} \\
    \hline
    DB-VirtualDispatch2 & $\true\falsepositive$ & $\true \truenegative$ & $\true \truenegative$\\
    DB-VirtualDispatch3 &  $\falsepositive$ & $\falsepositive$ & $\truenegative$\\
    DB-StaticInitialization1 &  $\falsenegative$  & $\true$  & $\true$ \\
    DB-StaticInitialization3 & $\falsenegative$ & $\true$  & $\true$ \\
    DB-StringFormatter1 & $\falsenegative$ & $\true$  & $\falsenegative$ \\
    DB-Serialization1 & $\falsenegative$ & $\true$  & $\falsenegative$ \\
    \hline
    \multicolumn{4}{|c|}{\textbf{Android Specific}} \\
    \hline
    DB-PrivateDataLeak3 &$\falsenegative$& $\true$ & $\falsenegative$\\
    DB-PublicAPIField2 & $\falsenegative$& $\true$ & $\true$ \\
    \hline
    \multicolumn{4}{|c|}{\textbf{Inter-Component Communication (ICC)}} \\
        \hline
             DB-ActivityCommunication2  &  $\true \truenegative \falsepositive$ &  $\true \truenegative \truenegative$ & $\falsenegative \truenegative \truenegative$\\
             DB-ActivityCommunication3&  $\true\falsenegative$ &  $\true \true$ & $\falsenegative \falsenegative$ \\
             DB-ActivityCommunication4  & $\falsenegative\true\truenegative\falsepositive$ &  $\true \true \truenegative \truenegative$ & $\falsenegative \falsenegative \truenegative \truenegative$ \\
            DB-ActivityCommunication5  &  $\true \true$     & $\true \true$  & $\falsenegative \falsenegative$\\
             DB-ActivityCommunication6  & $\true \falsenegative $ &  $\true \true$ & $\falsenegative \falsenegative $\\
            IB-PrivateDataLeak3   & $\falsenegative $ &  $\true $ & $\falsenegative $\\
    \hline

    \multicolumn{4}{|c|}{\textbf{Callbacks}} \\
    \hline
    DB-Button3  & $\true$ & $\true$  & $\falsenegative$ \\
    DB-MethodOverride1& $\true$ & $\true$ & $\true$ \\
    \hline 

    \multicolumn{4}{|c|}{\textbf{Threading}} \\
    \hline
    DB-AsyncTask1  & $\true$ & $\true$  & $\falsenegative$ \\
    DB-Executor1 & $\true$ & $\true$ & $\falsenegative$ \\
    \hline 

    \multicolumn{4}{|c|}{\textbf{Arrays and Lists}} \\
    \hline
    DB-ListAccess1  &$\falsepositive$& $\falsepositive$ & $\truenegative$ \\  
    DB-ArrayToString1 &$\falsepositive$& $\true$ & $\falsenegative$ \\  
    \hline
    \multicolumn{4}{|c|}{\textbf{Sum, Precision, Recall, and $F_1$ measure}} \\
    \hline
        True positives \#, TP   &  97         &    111       &  36 \\
        True negatives \#, TP    &  23         &    28      &  31 \\
        False positives \#, FP   & 11           & 6            & 3 \\
        False negatives \#, FN  & 14           & 0             &  75 \\
        Precision, $ p = \frac{\text{TP}}{\text{TP}+ \text{FP} }$    & 89.81\%               & 94.87\%          &  92.31\%  \\
        Recall, $ r = \frac{\text{TP}}{\text{TP}+ \text{FN} }$   & 87.39\%           & 100\%          & 32.43\% \\
        $F_1$ measure, $2pr/(p+r)$     & 0.89               & 0.97          & 0.48 \\

    \hline

    \end{tabular}
     }
\end{table}

\begin{table*}[t]
    \centering
    \caption{The matching results of our divide and conquer strategy and the pure backtracking search for the highlighted cases.}
    \label{table:algorithm_performance}
\resizebox{\textwidth}{!}{
    \begin{tabular}{|l|c|c|c|c|c|c|c|c|c|c|c|}  
        \hline
        \multirow{2}*{\textbf{App Package Name}} & \multirow{2}*{\textbf{\# of Matched Logs}}& \multirow{2}*{\textbf{\# of Nodes}} & \multirow{2}*{\textbf{\# of Branch Nodes}}   & \multicolumn{4}{c|}{\textbf{Our Divide and Conquer Strategy}} &  \multicolumn{2}{c|}{\textbf{Backtracking Search}} \\
        \cline{5-10}
        & & & & \textbf{Time (sec)} & \textbf{\# of Visited Nodes} & \textbf{\# of Backtracks} & \textbf{Correct?} & \textbf{Time (sec)} & \textbf{\# of Backtracks} \\
        \hline
        ynqgas.mqbgseos & 15713 & 2634 & 604 & 531.8 & 2236 & 96022 & $\surd$ & \textgreater 3600 & \textgreater 421920 \\
        \hline
        ngjvnpslnp.iplhmk & 3507 & 1830 &  953  & 42.3 & 260 & 70939 & $\surd$ & \textgreater 3600 & \textgreater 1382039 \\
        \hline
        com.android.system.admin & 1231 & 2371 &  1176 & 84.3 & 116 & 253823 & $\surd$ & \textgreater 3600 & \textgreater 2171368 \\
        \hline
    \end{tabular}}
\end{table*}

\subsubsection{Complmentations to Static Data-flow Analysis}

To validate the reliability of our log matching in handling different Android scenarios, we verify if the revealed path of \oursystem complements the data-flow detectability of FlowDroid \cite{DBLP:conf/pldi/ArztRFBBKTOM14}.
In comparison, we choose IccTA \cite{DBLP:conf/icse/0029BBKTARBOM15} to improve the accuracy of ICC-based data-flow tracking in FlowDroid.
Furthermore, we select an efficient hybrid analysis tool named AppAudit~\cite{DBLP:conf/sp/XiaGLQL15} to test the apps in micro-benchmarks.
AppAudit's implementations are unavailable, so we cannot complement its data-flow detectability directly but regard its analysis results as the reference to assess our complementations on FlowDroid.
It provides a website for uploading APK files and receiving analysis results.
Since AppAudit does not treat some sinks (e.g., \texttt{Log.d()}) as sensitive operations, to avoid reducing recall caused by this reason, we replace all the sinks in the code of testing apps with \texttt{sendTextMessage()} beforehand.

We review the apps' code and manually operate them on the real device to collect logs.
Then we combine the revealed path of each app with its inter-procedural control-flow graph (\ie, ICFG) built by FlowDroid to produce the complemented ICFG automatically.
The complementation is to rebuild the missed call relations of some APIs (\eg, for unresolved reflective calls), prune the call relations uncovered at runtime (\eg, for imprecise ICC links), and decide the method call sequence (\eg, for uncertain transitions in Android lifecycle) in the ICFGs.
We modify the implementation of FlowDroid to make it perform data-flow analysis on the original and complemented ICFG respectively.

The experimental results show that FlowDroid achieves the data-flow tracking for more apps in micro-benchmarks with the complementation of \oursystem.
The representative detection results are shown in \Cref{tab:comparision_flowdroid_iccta}, where each symbol represents a detection result for a data-flow path and the number of symbols in a row means the number of data-flow paths detected from an app.
In the table, precision and recall of FlowDroid increase from 89.81\% and 87.39\% to 94.87\% and 100\% respectively after the complementation.
Meanwhile, the precision of the complemented FlowDroid is higher than the precision of AppAudit.
Note that the results of FlowDroid + AppAngio and FlowDroid + IccTA + AppAngio are the same, because in the complemented ICFGs the precise ICC links of IccTA are preserved and the imprecise ICC links of IccTA are pruned according to the runtime information of \oursystem.

Furthermore, limited by the nature of static analysis, some cases (\eg, \textit{ListAccess1} or \textit{VirtualDispatch3}) still cannot be solved by FlowDroid even if the ICFG is complemented.
AppAudit uses the dynamic execution to check if the data leaks detected by static analysis can happen in real execution, so it produces less false positives.

\subsubsection{Case Studies}
By manual checking, we confirm that the ICFGs built for various scenarios of Android apps are all complemented correctly.
We use three representative cases to show the effectiveness of AppAngio in complementing ICFGs. 

\noindent \textbf{Deciding the Method Call Sequence.} In \textit{Button2}, the data leakage will occur only if the user clicks a button in a specific order.
We collect different log sequences corresponding to different UI operating orders to perform the log matching and complement the ICFG of the app according to the matched paths.
The results show that the data leakage is detected by FlowDroid on only one of the complemented ICFG, which validates that the call relations uncovered at runtime are precisely pruned from the original ICFG and the method call sequence is decided based on the sequence of log records.

\noindent \textbf{Amending the ICFG.} The result of \oursystem helps find a design flaw of FlowDroid in modeling the lifecycle.
Specifically, there is an app named \textit{MethodOveride1} in DroidBench.
FlowDroid models \code{onCreate()} as the first method invoked in the dummy-main method of the app.
Actually, the overridden method \code{attachBaseContext()} in this case is invoked before \code{onCreate()} at runtime.
The problem is corrected by combining the original ICFG with the revealed path, because the path indicates the real execution sequence of the methods.
This flaw may cause false negatives for detecting some sophisticated cases, though it does not affect the result of the data-flow analysis in \textit{MethodOveride1}.
Furthermore, malicious developers may use this flaw to elude the detection of FlowDroid deliberately.

\noindent \textbf{Differentiating Log Points.} The case of \textit{ImplicitFlow3} in DroidBench exemplifies the method of \oursystem in differentiating the same log points on the two branches of the same method.
There are two \texttt{Log.i()} in the method named \texttt{leakInformationBit()}.
It is insufficient to differentiate which log point was executed at runtime only by the signature of \texttt{Log.i()}.
As mentioned in \Cref{logginginformation}, our tool can be extended to log the arguments of APIs to differentiate log points.
Specifically, \oursystem distinguishes the two log points because the second arguments of them are different (\ie, one is "0" and the other is "1").
If the arguments of the logged APIs cannot be determined statically sometimes, \eg, the strings are preprocessed by encryption or code obfuscation, \oursystem supports logging more auxiliary information (\eg, the types of the used arguments or the type of the return value) to differentiate the branch paths.

\begin{table*}[t]
\centering
\caption{The comparison of information obtained from the reports of VirusTotal and from the matched paths of AppAngio for 18 apps (the first 10 apps are malware samples and others are market apps).}
\label{tab:contextualrecovery}
\resizebox{\textwidth}{!}{
\begin{tabular}{|l|l|l|}
\hline
\textbf{App Package Name}      & \textbf{Reports of VirusTotal}                                                                                                                                                             & \textbf{Information Revealed from the Matched Paths of AppAngio}             \\\hline
1. ynqgas.mqbgseos                         & \begin{tabular}[c]{@{}l@{}}1. 82 methods, \eg, \texttt{java.lang.String.toCharArray()}, are invoked \\ via reflection \\ 2. IMEI, IMSI, phone number, \etc are obtained \\ 3. Detailed IP address, HTTP requests, DNS resolutions fornetwork communication \\ are used
\end{tabular}  & \begin{tabular}[c]{@{}l@{}}1. Almost all methods in the path (\eg, \texttt{java.lang.String.toCharArray()}) \\ are invoked by the reflective API named \texttt{invoke()}\\ 2. Almost all method names, class names and strings in the path are obfuscated, e.g., strings are \\ obfuscated as \texttt{13GatVBbt3yV<b}, \texttt{Ba9@3R9aR9o+a7}, \etc\\ 3. There is a suspicious conditional statement that checks if each character of a string can \\ be transferred to a value with the integer type, which is explained in Section V-D1 \\ 4. The user’s IMEI and IMSI are obtained by the embedded library \texttt{gdadbjrj.tbhcwnn.*}\\ 5. Network connection is executed by the code of the host app \end{tabular}  \\\hline

2. ngjvnpslnp.iplhmk                 & \begin{tabular}[c]{@{}l@{}}1. 57 methods, \eg, \texttt{java.lang.System.arraycopy()}, are invoked \\ via reflection \\ 2. IMEI, IMSI, phone number, \etc are obtained \\ 3. Detailed IP address, HTTP requests, DNS resolutions for network communication \\ are used

\end{tabular}                         & \begin{tabular}[c]{@{}l@{}}1. Almost all methods in the path (\eg, \texttt{java.lang.System.arraycopy()}) \\ are invoked by the reflective API named \texttt{invoke()}\\ 2. Almost all method names, class names and strings in the path are obfuscated, e.g., class names are \\ obfuscated as \texttt{ngjvnpslnp.iplhmk.yqniqkxgpoo}, \texttt{ngjvnpslnp.iplhmk.gsnuxoavi}\\ 3. The user’s IMEI, IMSI and phone number are obtained by the code the host app\\4. Network connection is executed by the code of the host app

\end{tabular}                                                                                                                                                                                     \\\hline
3. com.android.system.admin         & \begin{tabular}[c]{@{}l@{}}
1. 117 methods, \\ \eg, \texttt{android.app.admin.DevicePolicyManager.isAdminActive()}, \\ are invoked via reflection \\ 2. Shell command (\ie, su -c 'id') is used \end{tabular}               & \begin{tabular}[c]{@{}l@{}}1. The API named \texttt{android.app.admin.DevicePolicyManager.isAdminActive()} \\ is invoked via reflection for checking if the given administrator component is active repeatedly\\ 2. Almost all methods names in the path are obfuscated, \eg, \texttt{oCIlCll()}, \texttt{OcIcoOlc()}\end{tabular}                                                                                                                                                                                                                                                                                      \\\hline
4. com.rbigsoft.easyunrar.wnvz   & \begin{tabular}[c]{@{}l@{}}
1. 13 methods, \eg, \texttt{com.cosc.k.IPCKUM.KU()} are invoked via reflection \\ 2. Detailed IP address, HTTP requests, DNS resolutionsfor network communication \\ are used \\
  3. File system actions, \eg, opening, deleting and writing files, are executed \end{tabular} & \begin{tabular}[c]{@{}l@{}}1. The APIs of the embedded libraries (\eg, \texttt{com.cosc}, \texttt{com.cosb}) are invoked via reflection\\ 2. Network connection is executed by the libraries named \texttt{com.cosb} and \texttt{com.ulk} \\3. The user’s operations on UI are recorded\end{tabular}                                                                                                                                                                                                                                                                                                                                                   \\\hline
5. org.baole.app.antismsspam     & \begin{tabular}[c]{@{}l@{}}
1. Location information and the ISO country code are obtained \\
2. File system actions, including opening, deleting and copying files, are executed
\end{tabular}                                    
& \begin{tabular}[c]{@{}l@{}}1. The user’s location information is collected by the advertising library \texttt{Admob}\\ 2. The user’s configurations, e.g., SPAM BLACKLIST, are recorded\\ 3. The password set by the user is saved\end{tabular}                                                                                                                                                                                                                                                                                            \\\hline
6. com.eamobile.sims3\_row\_qwf     & \begin{tabular}[c]{@{}l@{}}1. Location information, phone number, IMEI are obtained \\ 2. Detailed IP address, HTTP requests, DNS resolutions for network communication \\ are used
\end{tabular}                                               & \begin{tabular}[c]{@{}l@{}}1. The user’s IMEI and location information are collected by the library named \texttt{PushAds} \\ when the app launches \\ 2. Network connection is executed by the library named \texttt{com.av111236} \end{tabular}                                                                                                                                                                                                                                                                                                                                                                                                                             \\\hline
7. com.example.child      & \begin{tabular}[c]{@{}l@{}}1. IMSI, IMEI, Network information, \etc are obtained  \\ 2. 3 methods, \eg, \texttt{com.zdt.shell.ShellActivity.startAd1}, \\ are invoked via reflection  \\ 
3. Detailed IP address, HTTP requests, DNS resolutions for network communication \\ are used \end{tabular}      & \begin{tabular}[c]{@{}l@{}}1. The user’s information, \eg, uid, timing mode, is collected by the library named \texttt{jypush}\\ 2. The APIs of the advertising library named \texttt{ELM} are invoked via reflection\\3. Network connection is executed by the library named \texttt{ELM} \\4. The unicode strings are the encoding results of voice data\end{tabular}                                                                                                                                                                                                                                                             \\\hline
8. com.pianfang.book               & \begin{tabular}[c]{@{}l@{}}1. IMEI, IMSI, MAC address, \etc are obtained
\\ 2. Network communication to the advertising server is executed \end{tabular}                                   & \begin{tabular}[c]{@{}l@{}}1. The user’s IMEI, IMSI, MAC address and country code are collected by the advertising library \texttt{WAPS} \\ 2. Network connection is executed by the advertising library named \texttt{WAPS} \end{tabular}                                                                                                                                                                                                                                                                                                                                                                                                                              \\\hline
9. com.doaspx.Happy       & \begin{tabular}[c]{@{}l@{}}1. IMEI, IMSI, MAC address, \etc are obtained \\ 2. Network communication to the advertising server is executed \\ 3. File system actions, \eg, opening, writing and deleting files, are executed  \end{tabular}                                   & \begin{tabular}[c]{@{}l@{}}1. The user’s IMEI, IMSI, MAC address and country code are collected by the advertising library \texttt{WAPS}\\ 2. Network connection is executed by the advertising library named \texttt{WAPS}\\ 3. The user’s settings, \eg, bookmark, are saved\end{tabular}                                                                                                                                                                                                                                                                                                                                   \\\hline
10. com.evilsunflower.compass    & \begin{tabular}[c]{@{}l@{}}1. Network communication to the advertising server is executed
\end{tabular}                                                                             & \begin{tabular}[c]{@{}l@{}}1. The user’s IMEI and IMSI are collected by the advertising library named \texttt{Domob} \\2. Network connection is executed by the code of the host app\end{tabular}                                                                                                                                                                                                                                                                                                                                                                                                                                                          \\\hline
11. com.sentra.lowongan       & \begin{tabular}[c]{@{}l@{}}1. File system actions, including opening files, are executed \\ 2. Network communication to the Google server is executed \end{tabular}                                                                         & \begin{tabular}[c]{@{}l@{}}1. The user’s configurations, \eg, javascript enabled, zoom enabled, are saved\\ 2. Some method names are obfuscated as \texttt{a()}, \texttt{b()}, \etc \\ 3. The API in the Google advertising library is invoked via reflection\end{tabular}                                                                                                                                                                                                                                                                                \\\hline
12. com.appmk.book.main        & No report                                                                                                                                                                              & \begin{tabular}[c]{@{}l@{}}1. The user’s search history is saved\\ 2. The activities that can be performed for the given intent about\\ \texttt{com.google.android.apps.circles.platform.PlusOneActivity} are checked by\\ the Google advertising library\end{tabular}                                                                                                                                                                                                                                                                        \\\hline
13. com.kkd.folca           & No report                                                                                                                                                                              & 1. The information of taking photos, \eg, duration time, storage path, is saved                                                                                                                                                                                                                                                                                                                                                                                                                                                      \\\hline
14. aboard.and.koabacus         & \begin{tabular}[c]{@{}l@{}}1. Network communication to the advertising server is executed
\end{tabular}                                                                             & \begin{tabular}[c]{@{}l@{}}1. The user’s MAC address and configurations (\eg, Jar version) are collected by\\ the advertising library named \texttt{mocoplex} \\ 2. Network connection is executed by the library named \texttt{mocoplex}  \end{tabular}                                                                                                                                                                                                                                                                                                                                                                                                                    \\\hline
15. jotart.sltheme.minimal       & No report                                                                                                                                                                              & \begin{tabular}[c]{@{}l@{}}1. The number of activities that can be performed for the given intents about \\ \texttt{ginlemon.flowerpro}, \texttt{ginlemon.flowerpro.special} and \texttt{ginlemon.flowerfree} \\ is checked repeatedly to decide if the specified launcher is installed successfully\end{tabular}                                                                                                                                                                                                                                                                                                \\\hline
16. me.barkfor.findspy          & No report                                                                                                                                                                              & No sensitive behavior is found                                                                                                                                                                                                                                                                                                                                                                                                                                                                                                     \\\hline
17. sms.smslegal.free     & No report                                                                                                                                                                              & No sensitive behavior is found                                                                                                                                                                                                                                                                                                                                                                                                                                                                                                     \\\hline
18. ru.realision.abyss    & No report                                                                                                                                                                              & No sensitive behavior is found    \\\hline                                                                                                                                                                                                                                                                                                                                                                                                                                                                                                
\end{tabular}}
\end{table*}

\subsection{RQ3: Effectiveness of Log Matching and Contextual Reveal for Real-world Apps} 
\label{sub:performance_on_realword_apps}
To verify the effectiveness of \oursystem in the log matching and contextual reveal, we cooperate Monkey \cite{Monkey} and manual operations to execute apps and collect logs.
Specifically, we use Monkey to generate and input 4000 pseudo-random events for each app automatically, and perform manual operations to cover each app's functionalities as many as possible and help Monkey handle the circumstances like adding friends, login, \etc
We finally collect the logs of each app from the device and execute \textit{Matching Module}.
More time needs to be spent on manual efforts (more than 20 minutes for each app on average), including operating apps and analyzing the revealed contextual information, than on the automated testing or the log matching (less than 10 minutes for each app on average).
Therefore, the size of our test set is not as large as automatic tools~\cite{pan2017dark,DBLP:conf/ndss/RasthoferAB14}.
We randomly choose 800 real-world apps from our datasets, among which 400 apps are from AndroZoo and 400 apps are from MalGenome project and VirusShare.

\subsubsection{The Effectiveness of Our Divide and Conquer Strategy}
\oursystem achieves the log matching on all apps in our test set and meanwhile, the identified paths are ensured to be matched with runtime execution flows by manual checks of at least two Android experts.
To highlight the effectiveness of our strategy in the log matching, we choose the cases of three real-world apps from our test set in \Cref{table:algorithm_performance}.
These apps contain a large number of reflective calls, so almost all logged APIs are \texttt{invoke()}.
Therefore, the apps are representative to verify if the positions of the log points are precisely decided by our strategy.
Moreover, there is no work, to the best of our knowledge, focusing on the same problem of the log matching with us. 
To assess the benefits brought by our proposed scheme in reducing the number of backtracks, we compare our divide and conquer strategy with the pure backtracking search, which is a downgraded version of our strategy.

The experimental results validate that our strategy effectively avoids backtracking as much as possible in achieving the log matching.
As listed in \Cref{table:algorithm_performance}, our strategy achieves the log matching with 96022, 70939 and 253823 backtracks for the three apps respectively, while the backtracking search does not find the paths corresponding to the runtime execution flows when performing more than 421920, 1382039 and 2171368 backtracks for the apps respectively.
After further inspection, we notice that our backtracks are mainly produced by searching the CFGs of the methods with complex structures, \eg, successive exception handling.
Moreover, the supergraphs that need to be explored for three apps contain 2634, 1830 and 2371 nodes respectively, where there are 604, 953 and 1176 branch nodes.
Our strategy spends 531.8 seconds, 42.3 seconds and 84.3 seconds to find the matched paths, during which only 2236, 260 and 116 nodes are visited.
In comparison, the backtracking search does not report the correct paths when it has been running for more than 3600 seconds.
The algorithm may have to find the correct paths by exhaustively visiting the nodes in supergraphs in all possible sequences at worst.

\subsubsection{The Effectiveness of Revealing Contextual Information}
To validate the effectiveness of \oursystem in the contextual reveal, we make the comparative study between the contextual information extracted from the matched paths of AppAngio and the reports of VirusTotal \cite{VirusTotal}.
Specifically, we first invite two Android analysts to scrutinize and summarize the information about sensitive behaviors from the matched paths of randomly selected 18 apps (including 10 malware samples and 8 market apps) from our datasets by manual review.
Then we resort to VirusTotal to check the 18 apps and collect the detection results.
VirusTotal is a website that inspects the submitted mobile apps with over 70 antivirus scanners and provides detailed reports, \eg, the APIs called via reflection, obtained sensitive data, the operations on the network and filesystem, to users. 
In the reports, the item named \texttt{BEHAVIOR} lists the succinct descriptions for the detected behaviors of the apps.
We treat the descriptions as the baseline for evaluating the effectiveness of AppAngio in the contextual reveal.

The comparison of the results between VirusTotal and AppAngio is shown in \Cref{tab:contextualrecovery}.
The contextual information revealed by \oursystem is based on the runtime logs collected in limited time but VirusTotal performs the comprehensive analysis on the submitted apps, we focus on examining if AppAngio can provide effective complementations to the results of VirusTotal, instead of comparing if all the outputs of VirusTotal are matched with the results of \oursystem directly.
Specifically, for each app, we present all the information revealed from the matched path of AppAngio in the third column.
Then we extract the results corresponding to the aforementioned information from VirusTotal's reports and list the results in the second column.

\begin{figure}[t]
        \centering

        \includegraphics[width=3.2in]{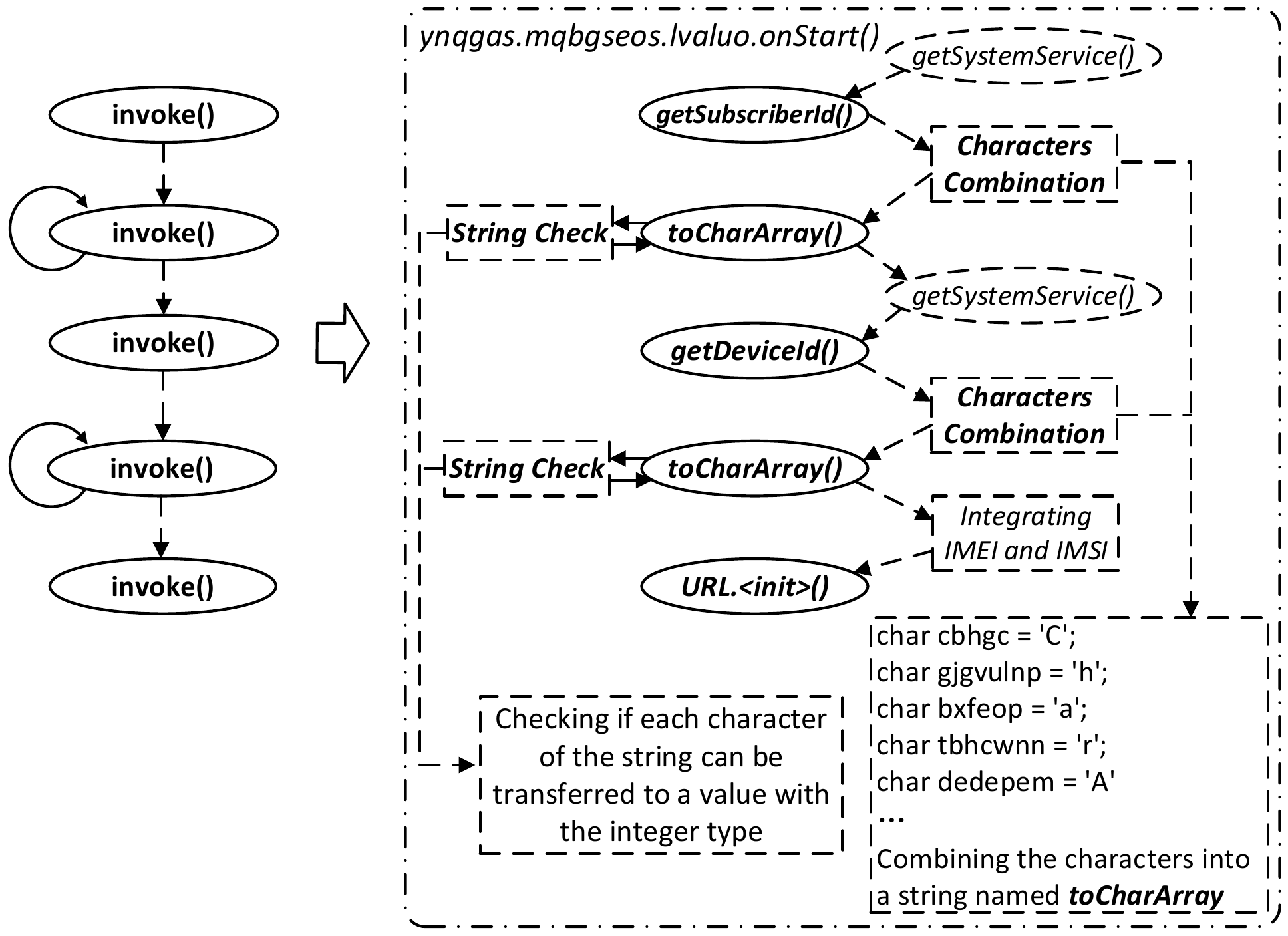}
        \caption{The comparison of two path segments built from the runtime logs and revealed by \oursystem respectively for the real-world app.}
        \label{fig:caseStudy}
\end{figure}

The comparative study manifests that the revealed details of \oursystem complement the reports of VirusTotal effectively.
Overall, all the details in the third column of \Cref{tab:contextualrecovery} enrich VirusTotal's reports from code obfuscation, user's operations, collectors of user's privacy, \etc
The enriched information can be complementary evidence for security analysts to inspect the internal logic of Android apps and capture potential malicious behaviors.
We exemplify the effectiveness of our work below:
\begin{enumerate}[1)]
\item \oursystem reveals that almost all the strings, method names and class names within the code of the app named \textit{ngjvnpslnp.iplhmk} are obfuscated, while VirusTotal does not report if the code of the app is obfuscated.
Based on the revealed contextual information, we have reason to doubt if the app is malicious.
\item \oursystem reveals that the app named \textit{com.doaspx.Happy} writes the user's settings, \eg, bookmark, display brightness, into a file, while VirusTotal only reports that the app performs the file writing operation.
We speculate that the operation is used to save the user's configurations.
\item \oursystem reveals that the app named \textit{com.eamobile.sims3\_row\_qwf} collects the user's IMEI and location information by the library named \texttt{PushAds}, while VirusTotal only reports that the app collects the user's private data.
According to the library name, we guess that the data collection operations are activated by the code in an ad library.

\end{enumerate}

\begin{figure}[t]
    \centering
    \includegraphics[width=0.95\linewidth]{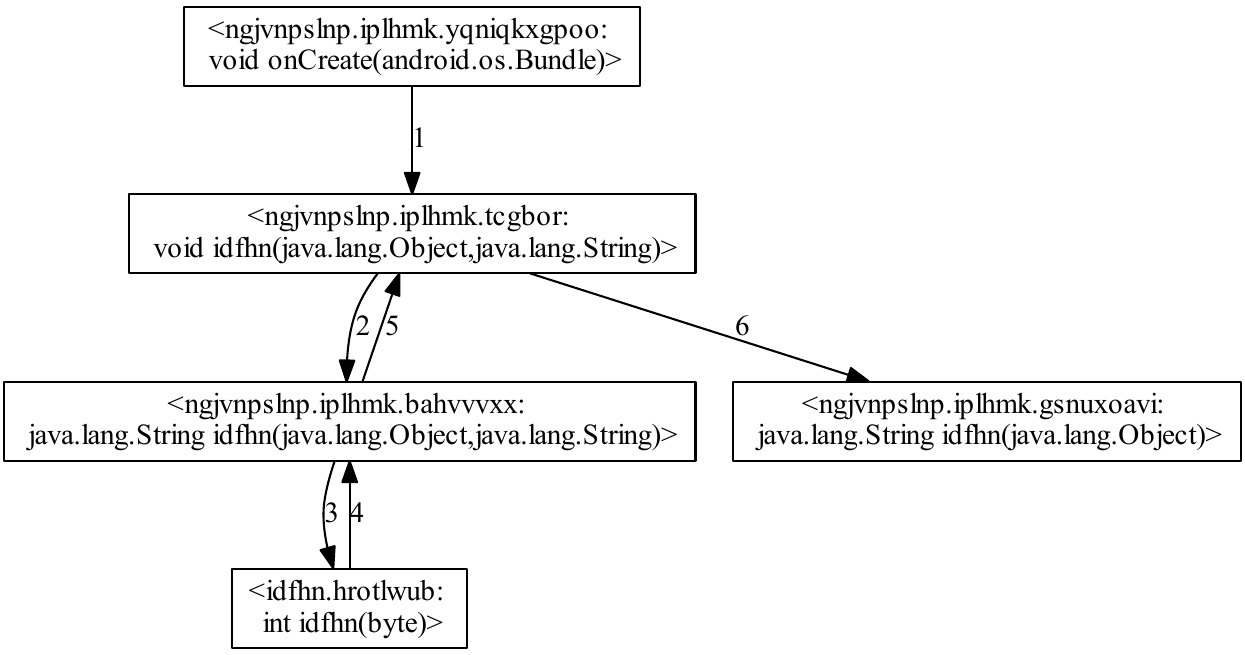}
    \caption{The method visit sequence matched with the given logs.}
    \label{fig:methodcall}
\end{figure}

\subsection{RQ4: Assistance in Identifying Maliciousness of Apps} 
\subsubsection{The Case of Disclosing Maliciousness within App Behaviors}
\label{subsubsec:case_study}
We present an app named \textit{gta3} as our case study.
It is a malware sample of FakeInstaller family, whose MD5 is \textit{dd40531493f53456c3b22ed0bf3e20ef} \cite{virusshare}.
In our evaluation, most apps only use the reflection mechanism occasionally. 
In this app, almost all the methods are invoked by reflective calls, and meanwhile, the arguments of the reflective calls are all obfuscated.
However, the signatures of reflective calls are not critical evidence to identify it as a malicious behavior due to the lack of contextual information.
Reflection techniques sometimes can be used for legitimate reasons, such as exploiting Android hidden and private APIs \cite{li2016droidra}.

The app is representative because it is challenging to obtain the real intention of its behavior.
On the one hand, it is impractical to find the path matched with runtime logs by the pure backtracking search because of the huge computational complexity.
Specifically, almost all the nodes about call statements in the app's supergraphs are \code{invoke()}.
Even when the app only runs about 20 seconds, \code{invoke()} is called for more than 11200 times.
The backtracking search cannot distinguish these log points in supergraphs.
On the other hand, traditional static analysis tools \cite{li2016droidra,DBLP:conf/pldi/ArztRFBBKTOM14} are inapplicable to reveal the real intention of the app's behaviors.
Harvester \cite{rasthofer2016harvesting} cannot extract all the runtime values for the app because of the limitation of the code slicing.
SherLog \cite{yuan2010sherlog} or lprof \cite{zhao2014lprof} do not apply to the case because the outputs of the log printing statements are inconsistent with the obfuscated argument strings in code.
AppAudit \cite{DBLP:conf/sp/XiaGLQL15} only finds that the app uses some sensitive permissions.
The existing behavioral reconstruction systems \cite{yan2012droidscope,DBLP:conf/ndss/TamKFC15,yuan2017droidforensics} may need to log sufficient actions of the app to reveal the required contextual information, which incurs considerable runtime overhead.

\oursystem reveals hidden maliciousness of the app successfully.
\Cref{fig:caseStudy} depicts two path segments, where the path on the left is built from the log records directly and the path on the right is the revealed result of \oursystem.
Obviously, the maliciousness of the app behavior is hard to be revealed based on the left result alone without any contextual information.
In comparison, with the right path, analysts can obtain a series of valuable information for assessing the behavior's maliciousness.
We list three highlighted findings as follows:

\begin{enumerate}[1)]
\item The embedded library collects user's privacy information (\ie, IMEI, IMSI) by calling \texttt{getDeviceId()} and \texttt{getSubscriberId()} when the app launches.

\item The app checks if each character of strings can be transferred to a value with the integer type.
When the check passes, the app transmits the strings to the specified server by the network.
We speculate that it is an anti-virtualization technique to avoid obtaining mock information \cite{rastogi2013droidchameleon}.

\item The app invokes \texttt{toCharArray()} by the reflection mechanism, where the method's signature is obtained by recombining a group of unordered characters, which may be a trick to evade static analysis.

\end{enumerate}

With the aforementioned contextual information, we have reason to conclude that this app behavior is malicious. 

\begin{figure}[t]
    \centering
    \includegraphics[width=\linewidth]{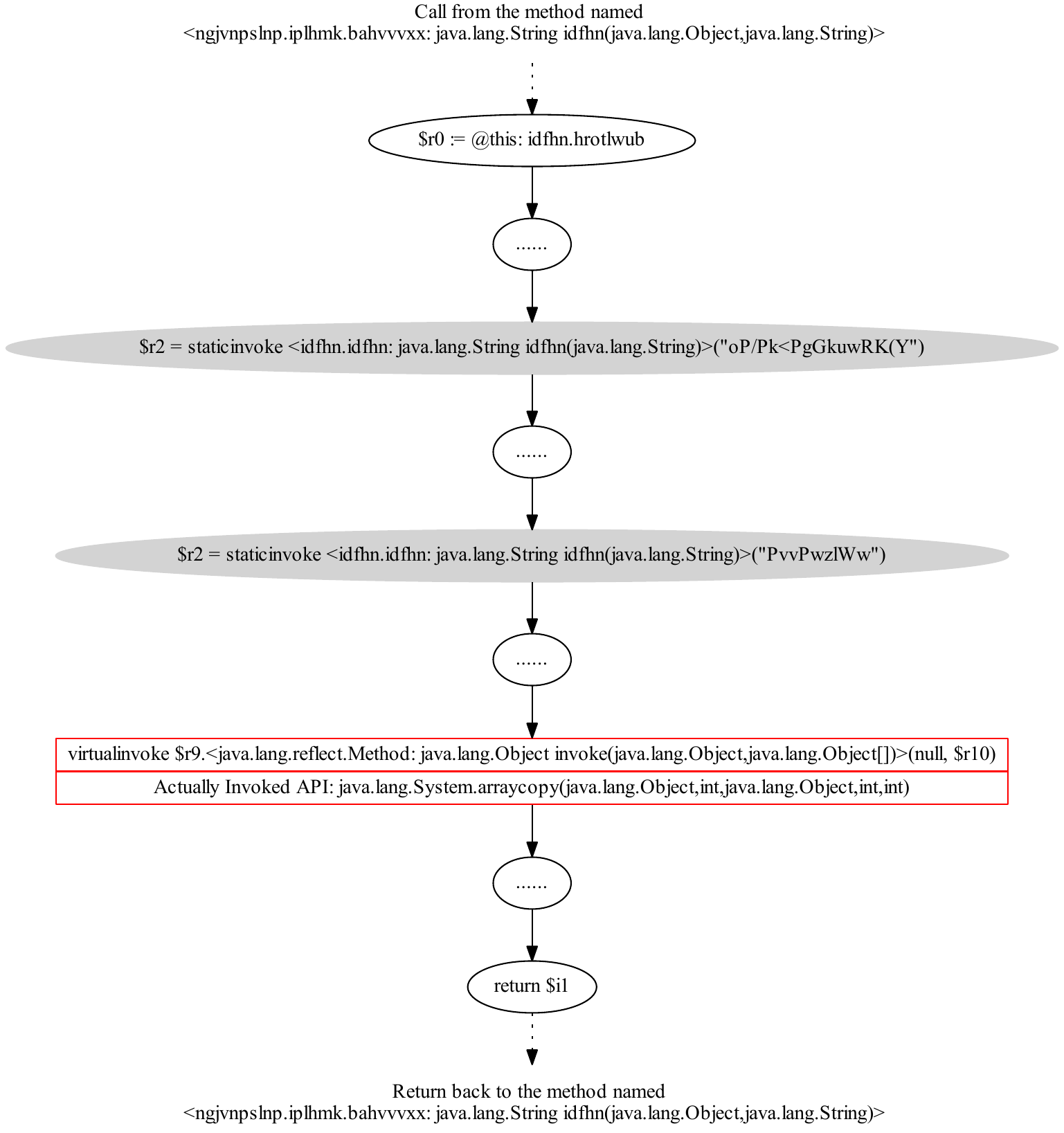}
    \caption{The simplified path segment corresponding to the method named \texttt{<idfhn.hrotlwub: int idfhn(byte)>}.}
    \label{fig:revealedMethod}
\end{figure}

\subsubsection{Ensurance for Manual Analysis}
To assist human experts in analyzing the matched path and retrieving critical information efficiently, we adopt the hierarchical mode to manage the path.
Specifically, AppAngio first builds a graph to depict the method visit sequence along the matched path.
It then supports displaying the path segment for a method or across multiple methods based on the configuration of analysts.
To display the graph and the path segment clearly, our tool automatically converts them to the related \texttt{DOT} files, which can be visualized by the existing tools (\eg, Graphviz).
In this way, the task of analyzing the matched path is divided into multiple sub-tasks, so the workload of manual analysis is relieved.

To exemplify the result of the above treatments, we extract 1000 log records about the app named \textit{ngjvnpslnp.iplhmk} to perform the log matching.
We exhibit the graph and the path segment in \Cref{fig:methodcall} and \Cref{fig:revealedMethod} respectively.
In \Cref{fig:methodcall}, the nodes represent method signatures and the edges labeled with integer numbers represent the visit sequence among the methods along the matched path.
\Cref{fig:revealedMethod} is the simplified path segment for the method named \texttt{<idfhn.hrotlwub: int idfhn(byte)>} in \Cref{fig:methodcall}.
Due to the page limit, the displayed path segment is simplified and some statements are represented as ellipses in nodes.
We create nodes with different styles to help analysts distinguish the elements in the graph.
For example, the upper layer of the red box is the log point in the app code (\ie, \texttt{invoke()}), and the bottom layer of the red box is the API that is actually invoked (\ie, \texttt{arraycopy()}).
The nodes filled with grey is the invocation statements where the callee methods are not related to any log points.
The dotted edges indicate the call relations for the method.
With the help of these indicators, analysts can extract the concerned information from the related path segments selectively, instead of analyzing the matched path directly.

 \begin{figure}[t]
    \centering
    \includegraphics[width=0.95\linewidth]{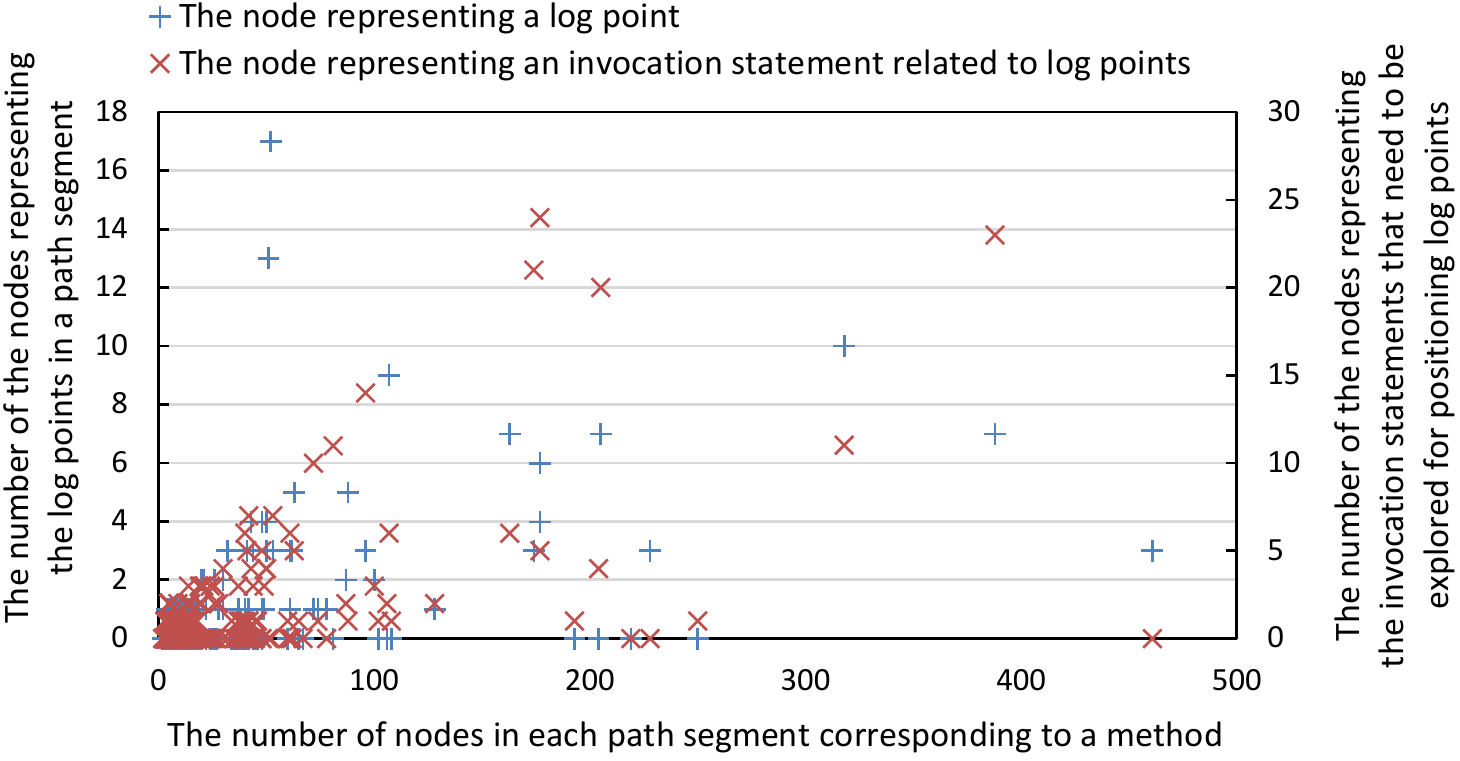}
    \caption{The statistics of the number of nodes within each path segment.}
    \label{fig:statistics}
\end{figure}

To validate the feasibility of the treatments, we calculate the number of nodes within the path segments of real-world apps, and the statistics are shown in \Cref{fig:statistics}.
The result demonstrates that the treatments ensure the efficiency of manual analysis and the workload of analysts is acceptable.
Specifically, \Cref{fig:statistics} is plotted based on 440 path segments, each of which corresponds to a method.
The x-axis is the number of nodes in each path segment (\eg, 128 nodes in \Cref{fig:revealedMethod}), the y-axis on the left is the number of the nodes representing the log points in each path segment (\eg, 1 of the nodes in \Cref{fig:revealedMethod}), and the y-axis on the right is the number of the nodes representing the invocation statements that need to be explored for positioning log points (\eg, \textit{V$_{2}$} in \Cref{fig:problemStatement}).
The average length of the path segments is 22, where 94.3\% of the segments include less than 3 log points and 92.2\% of the segments contain less than 3 invocation statements that need to be explored.
In other words, the number of key elements that need manual analysis in each path segment is limited.
Analysts can give priority to exploring a handful of the callee methods related to log points and focus on retrieving critical information around a small number of log points. 
\section{Limitations and Discussion}
\label{sec:limitations}

\oursystem achieves to help analysts reveal contextual information about app behaviors, but like any approach, it comes with some limitations.
We next point out the limitations and discuss possible solutions for security analysts in practice.

\subsection{The Trade-off between the Volume of Logged Data and the Effectiveness of Log Matching}\label{lim:trade-off}
Logging more information improves the effectiveness of the log matching but increases the performance overhead.
To balance the trade-off, \oursystem may output multiple paths matched with a given log segment with insufficient logging information.
Specifically, if all branches of a node involve the same sequence of logged APIs or do not involve any logged APIs, \oursystem may be unable to distinguish which branch is executed in real-time.
If the ambiguous cases are found, \oursystem presents them to analysts for making further identification and adjusting the existing logging strategy. 

In practice, security analysts can customize the logging configurations according to their requirements.
If the analysts plan to perform a coarse-grained and efficient analysis on app behaviors, they can log a small number of runtime information.
When they tend to make a fine-grained analysis, they can log more of the information.
In the future, we will try to design a machine learning-based approach for deciding the sweet-spot of the trade-off based on the features of different scenarios.

\subsection{Mechanisms for Disabling Our Work}

\oursystem may be evaded by anti-analysis techniques, \eg, code obfuscation, dynamic code generation.
As demonstrated before, \oursystem addresses a part of the
obfuscated code and then rebuilds the missed method call relation in the ICFGs.
However, there are various code obfuscation methods in practice, our solution is unable to address all the obfuscated cases in Android.
We plan to extend our work for solving more code obfuscation problems by integrating state-of-the-art deobfuscation techniques~\cite{rasthofer2016harvesting,mathis2017detecting} in the future.
Moreover, \oursystem does not support solving dynamic code generation currently.
In the future, we plan to extend our work to collect sufficient information for revealing app behaviors in the dynamically generated code with low performance overhead. 

Due to the limitation of the Soot framework, \oursystem does not support building the graph structure and performing the log matching based on native code temporarily, though it can log and present the APIs invoked via JNI.
After further investigation, we consider that the log matching process on the graph structure of native code is similar to the process on the graph built from Java code, where the critical idea of the log matching is identical. Therefore, in the future, we plan to combine our matching algorithm with the techniques that support modeling native code \cite{wei2018jn} to facilitate the identification of maliciousness of app behaviors within native code. 

\oursystem may be disabled by the kernel-level attacks.
Specifically, the attackers can tamper with the logs temporarily stored in the user's device by exploiting the kernel-level vulnerability.
This causes that no matched path can be found in an app's CFG.
The optional methods to achieve secure logging include access control, data encryption,~\etc
We believe that it is orthogonal to our primary purpose.

\subsection{Manual Efforts Required for Handling Revealed Results}
Similar to taint-flow analysis, the core procedure (\ie, recording runtime logs and identifying the path based on the CFG and the logs) of \oursystem is automated, while our tool still requires manual efforts to comprehend contextual information of the revealed path and gather their concerned evidence on the path to assess the app's maliciousness.
A possible solution is to design a machine learning-based system to achieve automated behavioral identification by extracting required contextual features from the matched path.
\section{Related Work}
\label{sec:related_work}

\subsection{Behavioral Reconstruction} 
\subsubsection{For the Android Platform} As mentioned before, many techniques have been proposed to reconstruct app behaviors via runtime information~\cite{DBLP:conf/ndss/TamKFC15,yan2012droidscope,yuan2017droidforensics,DBLP:conf/ccs/ZhangYXYGNWZ13}.
By combining the logs with CFGs, \oursystem reveals the path with the contextual information that is unavailable from the logs directly.
Moreover, the result of \oursystem may be cooperated with the techniques to achieve more accurate behavioral reconstruction.

\subsubsection{For Other Platforms} Besides the techniques for the Android ecosystem, some tools achieve behavioral reconstruction for other platforms.
SLEUTH~\cite{hossain2017sleuth} reconstructs the real-time attack scenario for the systems of Windows, Linux, and FreeBSD by audit-log data.
SherLog~\cite{yuan2010sherlog} helps programmers diagnose the errors of the software and systems with runtime logs.
lprof~\cite{zhao2014lprof} stitches runtime logs by analyzing the temporal relationships between log printing statements.

\oursystem faces different challenges with SherLog and lprof.
Specifically, SherLog matches the dynamically generated strings in logs with the source code, and lprof obtains the output string of each log printing statement by static analysis.
In the Android ecosystem, app source code is commonly unavailable for security analysts and there exist some anti-analysis techniques \cite{rasthofer2016harvesting,rastogi2013droidchameleon} (\eg, string encryption, code obfuscation) to hinder static or manual analysis.
Therefore, it is inapplicable to match the outputted logging messages with argument strings of log printing statements extracted from app code directly.
In \oursystem's implementation, we address the problem by leveraging runtime logs to resolve the strings in app code even if the strings are obfuscated.

\subsection{Behavioral Analysis} 
\subsubsection{Static Analysis}
The performances of existing static analysis techniques, \eg, signature-based schemes \cite{DBLP:conf/mobisys/GraceZZZJ12,feng2014apposcopy} or data-flow tracking schemes \cite{DBLP:conf/pldi/ArztRFBBKTOM14,DBLP:conf/icse/0029BBKTARBOM15,DBLP:conf/ccs/WeiROR14}, all rely on the completeness and precision of the underlying graph structures (\eg, callgraph or CFG).
The revealed paths of \oursystem can be used to circumvent some defects of static analysis and complement the existing graph structures.

\subsubsection{Dynamic Analysis} TaintDroid \cite{enck2014taintdroid} modifies the Dalvik virtual machine to implement the dynamic taint tracking.
AppsPlayground \cite{DBLP:conf/codaspy/RastogiCE13} adopts an improved version of TaintDroid for the dynamic data-flow tracking.
Due to different implementation goals, \oursystem does not insert any module inside the OS to track runtime data flows or perform the elaborate path exploration for testing apps; instead, it only logs the target API calls at runtime.
Therefore, the performance overhead imposed by \textit{Logging Module} of \oursystem is less than the techniques with heavyweight computations.

AppAngio adopts a standard logging technique that is similar to Quire \cite{dietz2011quire} and Compact \cite{wang2014compac}, while the implementation goals of the three tools are different, which makes that the tools process the logged information differently.
Specifically, AppAngio aims to reveal contextual information in Android app behaviors.
Therefore, it links the logged API callsites orderly on the app’s CFG based on the difference of the call stack information (\ie, \textit{Csi}) between successive log records.
In comparison, Quire aims to address the privilege escalation problems across different apps, so it leverages \textit{Csi} to reason about the provenance of sensitive operations before granting access to the requesting app.
Compac aims to achieve a fine-grained access control at app's component level, so it uses \textit{Csi} to extract Java package call chain at runtime.

\subsubsection{Hybrid Analysis}
The effective combination of static and dynamic analysis schemes can circumvent some of their respective shortcomings and obtain the required results more efficiently and precisely.
AppIntent \cite{DBLP:conf/ccs/YangYZGNW13} identifies possible execution paths about sensitive data transmission and lets human analysts determine if they are user-intended.
AppAudit~\cite{DBLP:conf/sp/XiaGLQL15} proposes an efficient analysis framework with less time and memory compared with AppIntent and FlowDroid.
Harvester~\cite{rasthofer2016harvesting} extracts the values of interest from apps.

AppAngio and the techniques solve the problems in different areas.
For instance, AppAngio achieves the log-based behavioral reconstruction and Harvester accomplishes the sensitive data extraction.
In the former area, AppAngio reveals abundant contextual information of the logged APIs and circumvents a part of the defects of static analysis in the graph building, while some of the contextual information may be unavailable for Harvester because of its limitations in the code slicing for ICC, modeling the Android lifecycle, \etc
In the latter area, Harvester supports obtaining the runtime values of the specified reflective calls in app code based on users' configurations, but \oursystem does not support it currently.
Therefore, the two techniques may complement each other by leveraging the technical advantages in their respective areas. 

\subsubsection{Machine Learning-based Analysis} 
Machine learning-based systems \cite{yang2015appcontext,DBLP:conf/icse/AvdiienkoKGZARB15} can identify malicious app behaviors. 
Imprecise graph structures caused by ICC, reflection or others bring difficulties in extracting precise features from app code.
For instance, the unresolved reflective APIs are treated as being security-sensitive~\cite{yang2015appcontext}, which causes false positives in detection.
The combination of \oursystem's results and the existing graph structures can help extract precise features for training more effective models for behavioral identification.
\section{Conclusion}
\label{sec:conclusions}
We propose and implement \oursystem, a novel system for revealing contextual information in Android app behaviors by API-level audit logs.
We design a divide and conquer strategy that uses call stack information at runtime to position each log point individually and guide the path exploration.
The experimental results on micro-benchmarks and real-world apps validate the precision and effectiveness of the log matching and the contextual reveal.
The matched paths of \oursystem help security analysts identify the maliciousness within real-world apps.
Meanwhile, \oursystem incurs negligible performance overhead (1.98\% on average) on the Android device.
\section*{Acknowledgements}
We thank the anonymous reviewers for their valuable comments.
The research is supported by the National Key R\&D Program of China 2018YFB2100300, 2018YFB0803400, National Natural Science Foundation of China under Grant No.61972369, No.61572453, No.61520106007, No.61572454, and the
Fundamental Research Funds for the Central Universities, No. WK2150110009.

\bibliographystyle{IEEEtran}
\bibliography{IEEEabrv, mybibfile}

\begin{IEEEbiography}[{\includegraphics[width=1in,height=1.25in,clip,keepaspectratio]{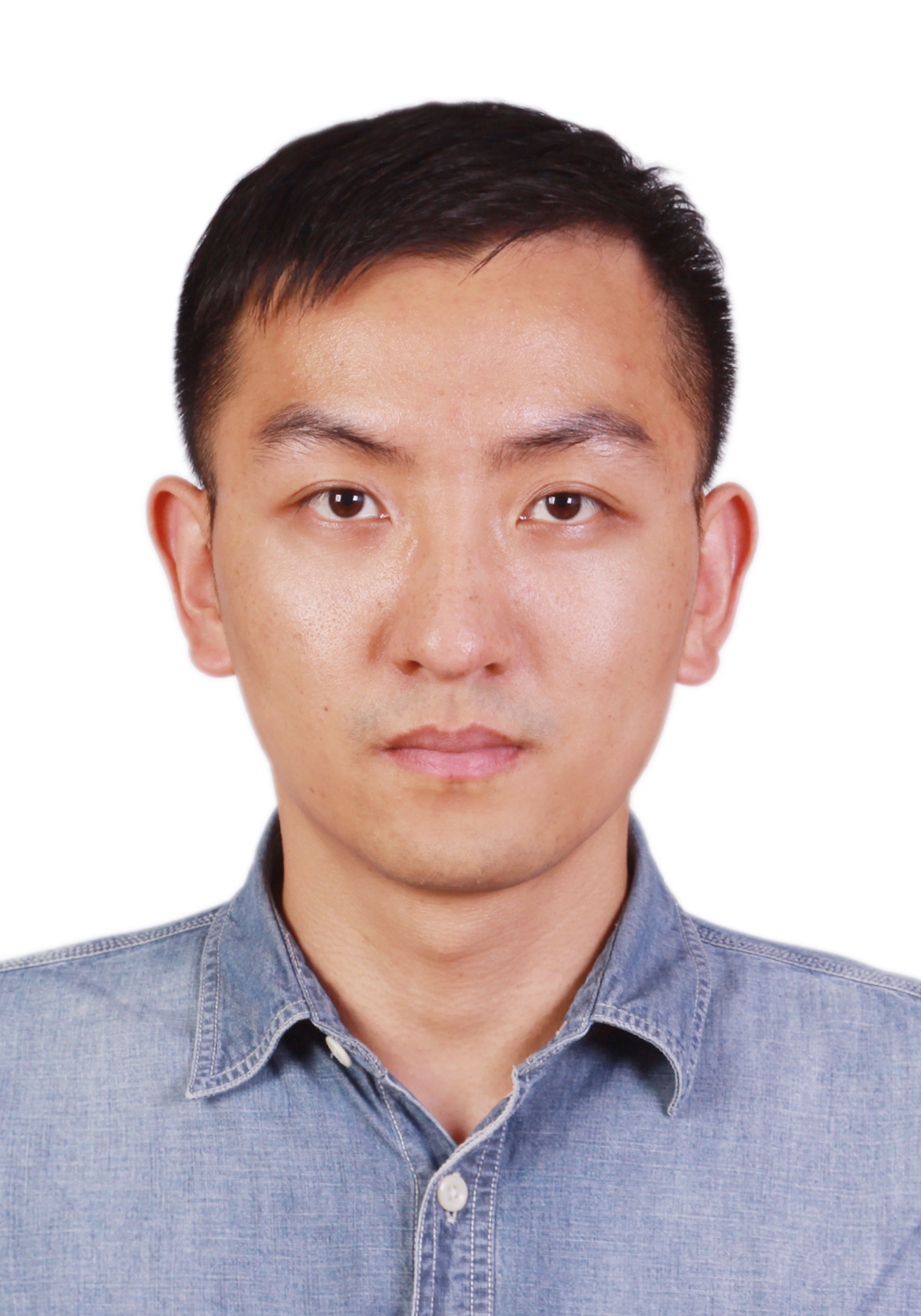}}]{Zhaoyi Meng} received the B.S. degree in information security from University of Electronic Science and Technology of China in 2014, and the Ph.D. degree in computer science and technology from University of Science and Technology of China. 
He is currently a Post-Doctoral Researcher with the Department of Computer Science and Technology, University of Science and Technology of China.
His current research interests include Android security and software formal verification.
\end{IEEEbiography}

\begin{IEEEbiography}[{\includegraphics[width=1in,height=1.25in,clip,keepaspectratio]{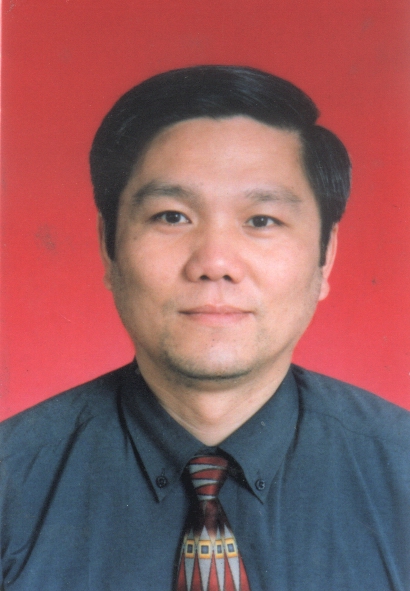}}]{Yan Xiong} received the B.S., M.S., and Ph.D. degrees from University of Science and Technology of China in 1983, 1986 and 1990 respectively. He is a professor in Department of Computer Science and Technology, University of Science and Technology of China. His main research interests include distributed processing, mobile computing, computer network and information security.
\end{IEEEbiography}

\begin{IEEEbiography}[{\includegraphics[width=1in,height=1.25in,clip,keepaspectratio]{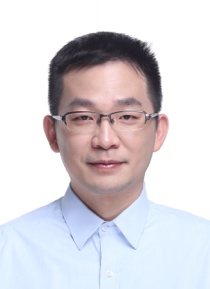}}]{Wenchao Huang} received B.S. and Ph.D. degrees from University of Science and Technology of China in 2006 and 2011 respectively. He is an associate professor currently with Department of Computer Science and Technology, University of Science and Technology of China. His current research interests include mobile computing, information security, trusted computing, and formal methods.
\end{IEEEbiography}

\begin{IEEEbiography}[{\includegraphics[width=1in,height=1.25in,clip,keepaspectratio]{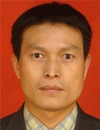}}]{Fuyou Miao} received the Ph.D. degree in computer science from University of Science and Technology of China, where he is an associate professor currently with Department of Computer Science and Technology. His research interests include information security, information coding key management in WSN, and network security.
\end{IEEEbiography}

\begin{IEEEbiography}[{\includegraphics[width=1in,height=1.25in,clip,keepaspectratio]{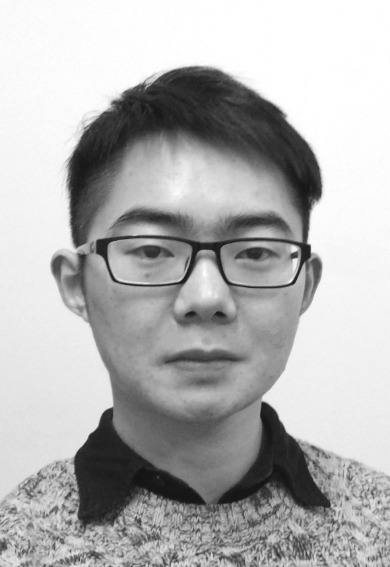}}]{Jianmeng Huang}
received the B.S. and Ph.D. degrees in computer science from University of Science and Technology of China in 2013 and 2018,
respectively. His current research interests include information security, mobile computing and IoT security.
\end{IEEEbiography}

\end{document}